\documentclass[aps,prl,twocolumn,groupedaddress,showpacs,floatfix,altaffilletter]{revtex4-1}
\usepackage{graphicx}
\usepackage{grffile}
\usepackage{amsmath}
\usepackage{amssymb}
\usepackage{bm}
\usepackage{color}
\usepackage[dvipsnames]{xcolor}
\usepackage{amsmath,amssymb,amsfonts}
\usepackage{epsfig}
\usepackage{times}
\usepackage[colorlinks,bookmarks=false,citecolor=blue,linkcolor=red,urlcolor=blue]{hyperref}

\newcommand{\fref}[1]{Fig.~\ref{#1}}
\newcommand{\eref}[1]{Eq.~(\ref{#1})}

\begin{document}

\title{Correlation-Enhanced Odd-Parity Interorbital Singlet Pairing in the Iron-Pnictide Superconductor LiFeAs}

\author{R.~Nourafkan$^{1}$, 
G.~Kotliar$^{2}$, 
and A.-M.S. Tremblay$^{1,3}$}
\affiliation{$^1$D{\'e}partement de Physique and Institut quantique, Universit{\'e} de Sherbrooke, Sherbrooke, Qu{\'e}bec, Canada J1K 2R1}
\affiliation{$^{2}$Department of Physics \& Astronomy, Rutgers University, Piscataway, NJ 08854-8019, USA}
\affiliation{$^3$Quantum Materials Program, Canadian Institute for Advanced Research, Toronto, Ontario,  M5G 1Z8, Canada}
%
%
\begin{abstract}
The rich variety of iron-based superconductors and their complex electronic structure lead to a wide range of possibilities for gap symmetry and pairing components. Here we solve  in the two-Fe Brillouin zone the full frequency-dependent linearized Eliashberg equations to investigate spin-fluctuations mediated Cooper pairing  for LiFeAs . The magnetic excitations are calculated with the random phase approximation on a correlated electronic structure obtained with density functional theory and dynamical mean field theory. The interaction between electrons through Hund's coupling promotes both the intra-orbital $d_{xz(yz)}$ and the inter-orbital magnetic susceptibility. As a consequence, the leading  pairing channel, conventional $s^{+-}$, acquires sizeable inter-orbital $d_{xy}-d_{xz(yz)}$ singlet pairing with odd parity under glide-plane symmetry. The combination of  intra- and inter-orbital components makes the results consistent with available experiments on the angular dependence of the gaps observed on the different Fermi surfaces. 
\end{abstract}

\maketitle
LiFeAs is a stoichiometric superconductor with superconducting $T_c\simeq18$~K and no magnetic ordering.~\cite{PhysRevB.78.060505} 
Despite rather poor nesting~\cite{Allan04052012, PhysRevLett.105.067002, PhysRevLett.109.177001,PhysRevB.83.060501}, recent quasiparticle interference experiments identify 
the antiferromagnetic (AF) spin-fluctuation mediated mechanism as the predominant pairing interaction.~\cite{Nat.Phys.11.177}  
ARPES and quasiparticle-scattering interference measurements below $T_c$ show that the superconducting (SC) gaps of LiFeAs are nodeless, with a Fermi surface (FS) dependence and a sizable variation along each FS.~\cite{PhysRevLett.108.037002,sym4010251, Allan04052012} 
Polarized neutron diffraction as a function of temperature has shown a suppression of the local spin susceptibility in the SC phase, suggesting singlet pairing.~\cite{PhysRevB.89.045141, PhysRevB.88.060401} 

In theoretical studies, the AF spin-fluctuation mediated pairing~\cite{PhysRevB.88.174516, PhysRevB.84.235121, PhysRevB.89.144513, Nat.Phys.10.845} and a combination of AF spin-fluctuation and orbital fluctuation mediated by phonons have been investigated.~\cite{PhysRevB.90.035104, doi:10.7566/JPSJ.83.043704}   
However, all studies are performed in the one-iron unit cell with various unfolding algorithms used to embed the correct symmetry.~\cite{PhysRevB.78.144517, PhysRevB.80.104503,  ANDP:ANDP201000149, PhysRevB.88.155125, 1367-2630-15-7-073006} 
This procedure is exact only for computing in-plane pairing.
In addition, the SC gap equation is usually projected on the FS, the pairing interaction is symmetrized,~\cite{PhysRevB.88.174516} and the resulting equation is always solved in the BCS approximation. All of the above simplifications must be questioned before we can be confident of the results. Furthermore, for Fe-based superconductors (FeSCs) with a non-symmorphic point-group,~\cite{PhysRevB.88.134510} anti-symmetry of fermions does not place a constraint on the parity of the SC pairing channel.~\cite{PhysRevX.3.031004,PhysRevB.89.045144} This allows for even-parity $d_{xz}-d_{yz}$ inter-orbital pairing~\cite{1410.3554}, or for $d_{xy}-d_{xz(yz)}$  odd parity spin singlet pairing when there is orbital weight at the Fermi level from orbitals with different in-plane mirror reflection symmetry~\cite{PhysRevLett.114.107002}.    
 
Hence, here we revisit spin-fluctuation mediated pairing by considering both Fe-$3d$ and As-$4p$ orbitals in the two-Fe unit cell.  We solve the linearized Eliashberg equations~\cite{SM} in the two-Fe Brillouin Zone (BZ) to investigate SC pairing and gap symmetry. Since there is increasing evidence that superconductivity does not emerge as a FS instability~\cite{Nat.Commun.6.7056}, we work in the orbital representation instead of projecting the gap equation on the FSs. Our results show that in the leading channel, with the conventional $s^{+-}$ symmetry, odd parity inter-orbital pairing accompanies the usual intra-orbital pairing and increases with interactions, in particular with Hund's coupling. In contrast to previous studies~\cite{sym4010251,PhysRevB.88.174516, PhysRevB.84.235121, PhysRevB.89.144513}  we find that this state can reproduce the angular dependence of the gap  on the electron pockets.

\begin{figure}
\begin{center}
\begin{tabular}{cc}
\includegraphics[width=0.48\linewidth]{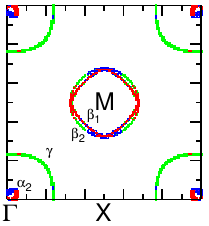} &
\includegraphics[width=0.48\linewidth]{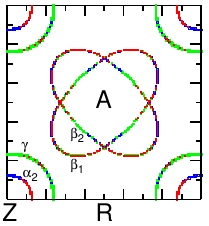}  
\end{tabular}
\end{center}
\caption{(Color online) Partial spectral weight, $A_{ll}({\bf k},0)$, of Fe $t_{2g}$- orbitals on the FS in the $k_x$-$k_y$ plane with $k_z=0$ (left), and  $k_z=\pi/c$ (right) obtained from the LDA+DMFT calculation. Here the $d_{xy}$, $d_{xz}$, and $d_{yz}$ orbitals are illustrated by green, blue and red colors, respectively. The $\alpha_1$ pocket crosses the Fermi level only in close vicinity to the $\Gamma$ point (not visible on this scale).}
\label{fig1}
\end{figure}

\paragraph{Electronic structure}
In LiFeAs, the bandwidth observed in ARPES is narrower than in LDA calculations and there are experimental evidences of long-lived magnetic moments.~\cite{PhysRevB.88.060401} This indicates the  importance of correlations, so we employ the LDA+DMFT method to obtain the electronic structure.~\cite{PhysRevB.81.195107, Nat.Mater.10.1038, PhysRevB.80.085101} 
 \fref{fig1} illustrates the LDA+DMFT partial spectral weight, $ A_{ll}({\bf k},0)$, of Fe $t_{2g}$- orbitals $d_{xy}$ and $d_{xz,yz}$ on the FSs of LiFeAs.~\cite{Note1}
 The Fe $e_g$ orbitals $d_{z^2}$ and $d_{x^2-y^2}$ hybridize with As-$p$ orbitals and contribute to the spectral weight lying above and below the Fermi level. 
 The FS consists of three hole-like and two electron-like sheets around the center and corners of the BZ respectively. The two inner hole pockets are predominantly composed from $d_{xz}$ and $d_{yz}$ orbitals.  The smallest hole pocket crosses the Fermi level only in close vicinity to the $\Gamma$ point. It hybridizes with the $d_{z^2}$ orbital near $Z$ point and is closed there, while remaining $2D$ away from this point. The middle pocket has 
moderate $k_z$ dispersion. The large hole-like Fermi surface originates purely from in-plane $d_{xy}$ orbitals and therefore is $2D$ without noticeable $k_z$ dispersion.  
The electron pockets are made from an admixture of  $d_{xy}$, $d_{xz}$ and $d_{yz}$ orbitals. The electron pockets intersect at small $k_z$ and their order flips, i.e., the inner pocket at $k_z=0$ is the outer pocket at $k_z=\pi/c$. 

Comparison to LDA,~\cite{SM} shows that in LDA+DMFT: (a) The two inner hole pockets shrink while the outer one expands. (b) The middle hole pocket also deforms and takes on a butterfly shape at small $k_z$.~\cite{PhysRevB.85.094505} (c) At finite $k_z$, the outer hole-pocket acquires some $d_{xz}$ and $d_{yz}$ orbital weight in the direction of the $A$ point. (d) The shrinkage of the two inner hole pockets leads to larger patches where $d_{xz}$ and $d_{yz}$ orbitals mix on these pockets. (e) The electron pockets are moderately expanded and they become closer to each other.~\cite{SM} 

The $t_{2g}$ orbitals are the most strongly correlated~\cite{Nat.Mater.10.1038, PhysRevB.85.094505} as is apparent from the mass enhancements $m^*/m_{LDA}=2.0$, $1.85$, $3.13$ and $2.7$ for $d_{z^2}$, $d_{x^2-y^2}$, $d_{xy}$, and $d_{xz,yz}$ orbitals, respectively. The $d_{xy}$ orbital has the strongest mass enhancement and shortest quasi-particle lifetime.

\begin{figure}
	\includegraphics[width=0.7\linewidth]{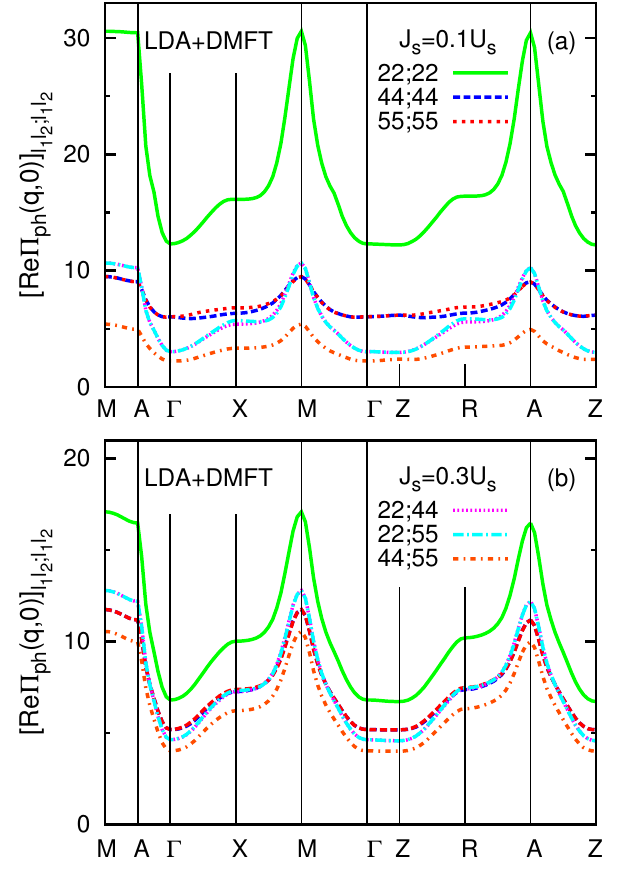}
	\caption{(Color online) Several components of the pairing interaction of LiFeAs at $k_BT=0.01$~eV in the particle-hole channel. 
		There are two sets of screened interaction parameters yielding the same magnetic Stoner factor, namely $J_s=0.1U_s$, $U_s=2.4$~eV on the top and $J_s=0.3U_s$, $U_s=1.68$~eV on the bottom. 
The legend for the color coding is spread over both figures. 
}\label{fig2}
\end{figure}

\paragraph{Effective pairing interaction}
A SC instability in the singlet channel occurs when the corresponding pairing susceptibility diverges as one lowers temperature. A divergent susceptibility signals the appearance of a pole in the corresponding reducible complex vertex function, which  describes all scattering processes of two propagating particles. 
Using the Bethe-Salpeter equation, the condition for an instability is that an eignvalue of the matrix $-{\bm \Gamma}^{irr, s}{\bm \chi}^{0}_{pp}$ becomes unity. Here ${\bm \Gamma}^{irr, s}$ is the irreducible vertex function (effective pairing interaction) in the singlet channel, and ${\bm \chi}^{0}_{pp}$ is the bare susceptibility in the particle-particle (p-p) channel.~\cite{Bickers2004, PhysRevB.57.5376, SM} 

The density/magnetic fluctuations contribute to the pairing interaction by entering the ladder vertex defined by ${\bm\Pi}_{ph}\equiv-(1/2){\bm\Gamma}^{irr, d}{\bm\chi}_{ph}^d{\bm\Gamma}^{irr, d}+(3/2){\bm\Gamma}^{irr, m}{\bm\chi}_{ph}^m{\bm\Gamma}^{irr, m}$ where ${\bm\chi}_{ph}^{m(d)}$ and ${\bm\Gamma}^{irr, m(d)}$ denote respectively the dressed susceptibility and the irreducible vertex function in the magnetic (density) channel.~\cite{SM} These vertices can be calculated in the DMFT approximation.~\cite{PhysRevB.86.064411} However, 
such a calculation is prohibitively difficult for multiorbital systems at the low temperatures necessary to study superconductivity,~\cite{SM} hence here we employ the random phase approximation (RPA).~\cite{PhysRevB.75.134519} In RPA, the irreducible vertex function is replaced by a static effective vertex which is parametrized by the \emph{screened} intra-orbital Hubbard interaction, $U_s$, and the Hund's coupling $J_s$.~\cite{SM, doi:10.7566/JPSJ.83.043704, PhysRevB.81.054518,PhysRevB.87.045113} The inter-orbital interaction and pair hopping are determined assuming spin-rotational symmetry.  Note that even though the static effective vertices $U_s$ and $J_s$ capture Kanamori-Br\"uckner screening effects, they do not fully capture the dynamics of screening. In particular, the RPA treatment misses the fact that at high fermionic frequencies one should recover the bare interactions. 

\fref{fig2} shows the pairing interaction, ${\bm\Pi}_{ph}$, at $k_BT=0.01$~eV for two sets of screened interaction parameters that yield the same magnetic Stoner factor.~\cite{Note2} Here we only present the intra-sublattice components because the inter-sublattice components are relatively small. In what follows, we focus on the Fe-$1$ and Fe-$2$ (on $A$ and $B$ sublattices respectively) $t_{2g}$ orbitals: $d_{xy}$ will be referred as $2$~($7$) and $d_{xz}$ and $d_{yz}$ orbitals as $4$~($9$) and $5$~($10$). The dominant effective pairing interaction components are repulsive. As can be seen in \fref{fig2}(a), due to better nesting, the $d_{xy}$ intra-orbital $(22;22)$ pairing vertex is dominant and the $d_{xz(yz)}$ intra-orbital $(44;44)$ is sub-dominant, yet on average it is larger than inter-orbital vertices $(22;44)$ and $(44;55)$. 

However, at larger $J_s/U_s$ the situation changes. For a fixed Stoner factor (proximity to magnetic transition) upon increasing the $J_s/U_s$ ratio from \fref{fig2}(a) to \fref{fig2}(b), the $d_{xy}$ intra-orbital pairing component decreases while the $d_{xz(yz)}$ intra-orbital components and the inter-orbital components increase slightly. This shows that  a higher $J_s$, through coupling to the more correlated $d_{xy}$ orbital, compensates the decrease of spin susceptibility expected from the lower $U_s$  (\fref{fig2}(b)).~\cite{SM} Furthermore, since Hund's coupling correlates different orbitals, the inter-orbital components increase, becoming comparable with the $d_{xz(yz)}$ intra-orbital components. The $d_{xy}$ intra-orbital vertex becomes less dominant at larger $J_s/U_s$.~\cite{Note3} This behavior of the magnetic susceptibility reflects itself directly in the pairing interaction (see supplemental material for the dressed susceptibilities in magnetic and charge channels). 

\begin{figure}
\includegraphics[width=0.7\linewidth]{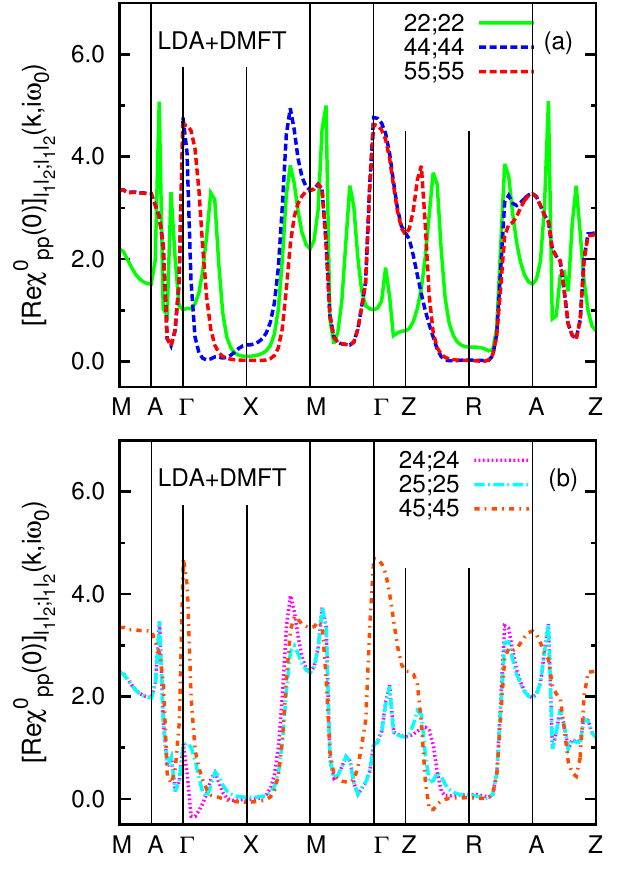}
\caption{(Color online) Real part  of the several intra-sublattice  components of the generalized particle-particle bare susceptibility at the lowest fermionic/bosonic Matsubara frequency. 
 }\label{fig3}
\end{figure}

\paragraph{Bare particle-particle susceptibility}
The generalized bare susceptibility in the p-p channel also enters the gap equation.~\cite{SM} \fref{fig3} shows the real part of several components of the generalized p-p bare susceptibility at the lowest fermionic/bosonic frequencies. The intra-orbital components are purely real. Both real and imaginary parts (see SM) show peaks at the position of FSs.  For example, going from the $\Gamma$ to the $X$ point in the top panel, the three peaks are respectively related to the inner hole pocket with $d_{xz}$ weight in close proximity to $\Gamma$, the middle pocket with $d_{yz}$ weight and the outer pocket with $d_{xy}$ weight. 
The peak heights are directly proportional to the corresponding orbital weight on the FSs and inversely proportional to the Fermi velocity. The peak widths are induced by correlation effects, implying that electrons near FSs may contribute to the Cooper pairing. In a non-interacting system the peak widths go to zero at zero temperature.~\cite{PhysRevB.93.241116} The larger $22;22$ peak component in the $M-\Gamma$ direction, compared with the $M-X(Y)$ direction, indicates that the SC gap on the outer electron pocket is larger in the $M-\Gamma$ direction. 

In the BCS approximation, only real parts survive for the components considered here, due to a summation over Matsubara frequencies. In this case, the inter-orbital pairing is suppressed. Including the imaginary part in the full gap equation changes this trend. The imaginary parts of the inter-orbital components change sign between corner and center of the BZ.  
They have some symmetries that transfer to the gap function: (i) They are odd under exchange of orbital indices, (ii) There is also a $\pi$ phase difference between the two Fe ions (see SM).

\begin{figure}
	\begin{center}
		\begin{tabular}{cc}
			\includegraphics[width=0.5\columnwidth]{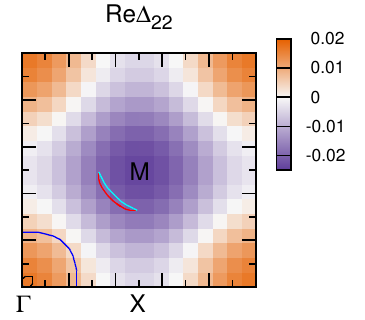} & 
			\includegraphics[width=0.5\columnwidth]{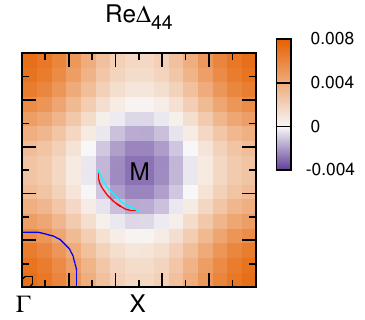} \\
			\includegraphics[width=0.5\columnwidth]{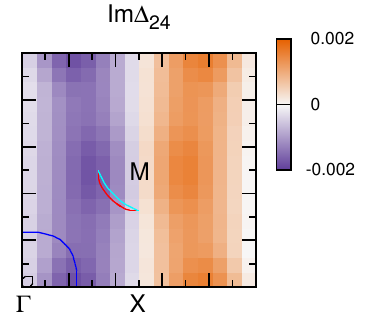} & 
			\includegraphics[width=0.5\columnwidth]{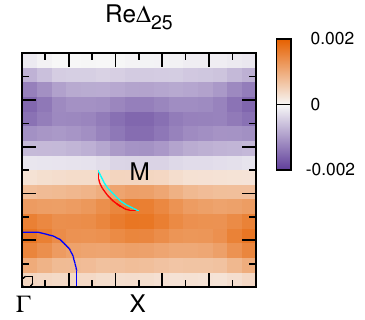} 
		\end{tabular}
		\caption{(Color online) Top panels: The real part of the $d_{xy}$ (left) and $d_{xz}$ (right) in-plane intra-orbital components of the SC gap function at the lowest Matsubara frequency with largest eigenvalue  in the orbital representation for $J_s/U_s=0.3$ and $k_BT=0.01$~eV. 
			The imaginary part can be obtained from ${\rm Im}  {\rm \Delta}_{ll}= -{\rm Re} {\rm \Delta}_{ll}$. Bottom panels: The real/imaginary part of the inter-orbital components of the SC gap function on sublattice $A$ in the orbital representation, $\Delta^{AA}_{l_1l_2}$. The corresponding components on sublattice $B$ are out of phase with the displayed components, i.e. $\Delta^{BB}_{l_1l_2}=-\Delta^{AA}_{l_1l_2}$. The lines show one quarter of the Fermi surfaces. }\label{fig4}
	\end{center}
\end{figure}

\paragraph{SC pairing symmetry in LDA+DMFT+RPA} 
The leading pairing channel is a channel with dominant $d_{xy}$, $d_{xz}$ and $d_{yz}$ intra-orbital pairing. In our gauge, the gap function components have both real and imaginary part which satisfy ${\rm Re}\Delta^{AA(BB)}_{ll} = -{\rm Im}\Delta^{AA(BB)}_{ll}$.  All intra-orbital components change sign between center and corner of the BZ (see \fref{fig4}), as expected in conventional $s^{+-}$ pairing. The $d_{xy}$ intra-orbital component dominates, but has a small value on the $\gamma$ pocket.  The $d_{xz}$ and $d_{yz}$ intra-orbital components are out of phase, i.e. $\Delta^{AA(BB)}_{55} \simeq -\Delta^{AA(BB)}_{44}$ (not shown). They take large values on the $\alpha_{1,2}$ hole pockets. 
The inter-sublattice components are much smaller than intra-sublattice ones, ${\bm \Delta}^{AA(BB)} >> {\bm \Delta}^{AB(BA)}$. The largest inter-sublattice component is ${\Delta}_{22}^{AB}$. In orbital basis, the gap functions do not change much between $k_z=0$ and $k_z=\pi/c$, hence we present only $k_z=0$ results.

In agreement with the above pairing-interaction analysis, upon increasing $J_s/U_s$ the $d_{xz/yz}$ intra-orbital pairing strengthen.  
Furthermore, the $d_{xy}$-$d_{xz}$ and $d_{xy}$-$d_{yz}$ inter-orbital pairings increase. Although they vary on a smaller interval, they are comparable with the $d_{xz/yz}$ intra-orbital components on the electron FSs (compare \fref{fig4} top and bottom panels). 

We verify that the gap function components of the leading channel 	satisfy the relations $\Delta^{AA(BB)}_{l_1l_2}({\bf k},i\omega_m)=\Delta^{BB(AA)}_{l_1l_2}(-{\bf k},i\omega_m)$, 
 and $\Delta^{AA(BB)}_{l_1l_2}({\bf k},i\omega_m)=\Delta^{AA(BB)}_{l_2l_1}(-{\bf k},-i\omega_m)$.~\cite{Note6}
The first relation says that the superconducting state does not break parity: In LiFeAs the inversion center is located in the middle of Fe-Fe link. Under parity operation the sublattice $A$ maps to sublattice $B$ and vice versa and ${\bf k} \rightarrow -{\bf k}$. The components of the gap function also satisfy the relation $\Delta^{AA(BB)}_{l_1l_2}(k_x,k_y,i\omega_m)=p_{l_1}p_{l_2}\Delta^{BB(AA)}_{l_1l_2}(k_x,k_y,i\omega_m)$, where $p_l$ denotes the parity of orbital $l$ with respect to in-plane mirror reflection symmetry.~\cite{Note7} This symmetry is defined by in-plane mirror reflection followed by a half-translation, expressed in units of the two-Fe unit cell, $\{\sigma^z|\frac{1}{2}\frac{1}{2}0\}$. Thus, the intra-orbital components on the two Fe are equal, 
while 
the inter-orbital components between one even-parity ($d_{xy}$) and one odd-parity ($d_{xz}$, $d_{yz}$) orbital, change sign between two Fe-ions. These components are the parity-odd under  $\{\sigma^z|\frac{1}{2}\frac{1}{2}0\}$ spin singlet pairings.~\cite{PhysRevLett.114.107002} 
Furthermore, as can be seen from  \fref{fig4}, the in-plane intra-orbital components satisfy $\Delta_{ll}^{AA(BB)}(k_x,k_y)=\Delta_{ll}^{AA(BB)}(-k_x,-k_y)$, while the inter-orbital components between $d_{xy}$ and $d_{xz(yz)}$ satisfy  $\Delta_{l_1l_2}^{AA(BB)}(k_x,k_y)=-\Delta_{l_1l_2}^{AA(BB)}(- k_x, k_y)$ or $\Delta_{l_1l_2}^{AA(BB)}(k_x,k_y)=-\Delta_{l_1l_2}^{AA(BB)}(k_x, - k_y)$ .

Our calculations show that the gap symmetry of the leading channel is conventional $s^{+-}$. Indeed, although there is a phase difference between the $d_{xz}$ and $d_{yz}$ components of the gap function in the orbital basis,  this phase difference is removed by another phase difference that arises when going to the Bloch basis corresponding to the $\alpha_{1,2}$ pockets.~\cite{SM} In the \emph{subleading} pairing channel, the $d_{xy}$ intra-orbital component is in phase with $d_{yz}$ and out of phase with $d_{xz}$ intra-orbital components, which in the band representation gives $s^{+-}$ gap symmetry with a sign change between $\alpha_{1,2}$ and $\gamma$ pockets and between electron pockets and accidental nodes on the $\beta_2$ pocket.~\cite{Nat.Phys.10.845}

\begin{figure}
	\begin{center}
		\begin{tabular}{cc}
			\includegraphics[width=0.8\linewidth]{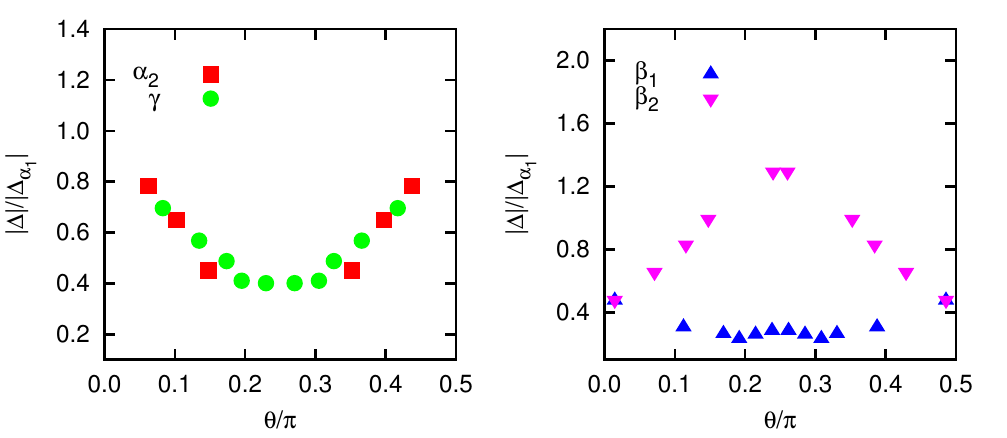} 
		\end{tabular}
	\end{center}
	\caption{(Color online) For $J_s/U_s=0.3$, the SC gap magnitude (in units of the average gap magnitude on the $\alpha{_1}$ pocket) as a function of the angle $\theta$  measured at the $\Gamma$ and $M$ points with respect to the $x$ axis for $k_z=0$ FSs. }\label{fig6}
\end{figure}

Finally, we comment on the SC gap magnitude on different FSs.~\cite{Note8} 
Diagonalizing the Bogoliubov quasi-particle Hamiltonian leads to a gap magnitude which has predominant $\cos4\theta$ angular dependence on all pockets, as can be seen from \fref{fig6}.
The angular dependence of the gap on the $\gamma$ and of the average gap on the $\beta_{1,2}$ pockets are consistent with ARPES data: The gap is maximum at $\theta=0,\pi/2$ and decreases when approaching $\theta=\pi/4$ (the direction toward $M$-point) on the $\gamma$ pocket, while the average gap is maximum at $\theta=\pi/4$ (direction toward $\Gamma$-point) on the $\beta$ pockets and decreases when approaching $\theta=0,\pi/2$ where the two pockets cross. The gap on the $\beta_2$ electron pocket is increased in the direction of $\Gamma$-point due to a larger $d_{xy}$ orbital content with a large pairing amplitude (see \fref{fig4}, upper panels). The gap on the $\beta_1$ electron pocket also shows a local enhancement at $\theta=\pi/4$. Due to interchange of electron pockets as a function of $k_z$, the gap on the inner pocket becomes larger than that on the outer pocket at a finite $k_z$. Hence, for these pockets, a direct comparison with ARPES data has to take averaging over a range of $k_z$ into account.~\cite{1409.8669}
The ratio between the average gap magnitude on $\beta$ pockets and $\gamma$ pocket is also consistent with ARPES results~\cite{PhysRevLett.108.037002, sym4010251}. 
However, the gap magnitude on the $\alpha$ pockets is not the largest. This discrepancy with ARPES results may come from the fact that ARPES is performed at very low temperature while the linearized Eliashberg gap equation is valid at temperatures infinitesimally close to the transition temperature. The tunneling spectroscopy study of LiFeAs has shown a temperature evolution of superconductivity.~\cite{1509.03431} A calculation at a lower temperature shows that the sharp peaks in the $44$ and $55$ bare paring susceptibilities, \fref{fig3}(a), grow  faster than the wider peak for $22$. This leads to an increase of the gap on the $\alpha$ pockets at lower temperatures.

\paragraph{Conclusion}
Solving the full linearized Eliashberg gap equation with both real and imaginary parts and including correlations in the LDA+DMFT framework leads to a detailed description of the leading pairing channel in LiFeAs. Accounting for correlations in the spin fluctuation approach allows to  correctly capture not only nesting effects but also Fe-$d$ orbital fluctuating moments with orbitally dependent dynamics. Although the intra-orbital $d_{xy}$ spin susceptibility is dominant, Hund's coupling between orbitals on individual Fe atoms promotes both the intra-orbital $d_{xz(yz)}$ component and the inter-orbital $d_{xy}$-$d_{xz(yz)}$ components of the magnetic susceptibility. As a consequence, the leading paring channel,  conventional $s^{+-}$, acquires inter-orbital singlet pairing component with odd parity under glide-plane symmetry. This type of pairing may also be realized in other iron-based superconductors.
Antiphase $s^{+-}$ pairing~\cite{Nat.Phys.10.845} is sub-leading. The combination of inter-orbital odd-parity and intra-orbital even parity singlet pairing leads to a description of the angle-dependence and of the relative magnitudes of the gap on the $\beta$ and $\gamma$ Fermi surfaces that is consistent with state of the art experiments.    

\begin{acknowledgments}
R.~N is deeply indebted to M.~E.~Pezzoli and F.~Marsiglio for many insightful discussions. We thank K.~Haule for his LDA+DMFT code and for discussions. R.~N and A.-M.S.~T are supported by the Natural Sciences and Engineering Research Council of Canada (NSERC) under grant RGPIN-2014-04584, and by the Tier I Canada Research Chair Program (A.-M.S.T.). G.~K is supported by the NSF-DMR1308141. We acknowledge the hospitality of the CIFAR quantum materials program. Simulations were performed on computers provided by CFI, MELS, Calcul Qu\'ebec and Compute Canada.
\end{acknowledgments}
 

%

\setcounter{equation}{0}
\setcounter{figure}{0}
\setcounter{table}{0}
\setcounter{page}{1}
\makeatletter
\renewcommand{\theequation}{S\arabic{equation}}
\renewcommand{\thefigure}{S\arabic{figure}}
\renewcommand{\bibnumfmt}[1]{[S#1]}
\renewcommand{\citenumfont}[1]{S#1}
\pagebreak
\begin{center}
\textbf{\large Supplemental Materials: Correlation-Enhanced Odd-Parity Interorbital Singlet Pairing in the Iron-Pnictide Superconductor LiFeAs}
\end{center}
Here, we present some details on the electronic structure of the LiFeAs in the first section. The Eliashberg equation and details of the calculation for the pairing interaction in the random phase approximation (RPA) and in the dynamical mean field theory (DMFT) are presented in the second section. We show how this calculation differs from the BCS approximation to these equations. Our results on the bare and dressed susceptibilities are presented in fourth section. Finally, we discuss some symmetries of the Green's function and their consequences on the susceptibilities.

\section{Electronic structure}
\begin{figure}
\begin{center}
\includegraphics[width=0.5\columnwidth]{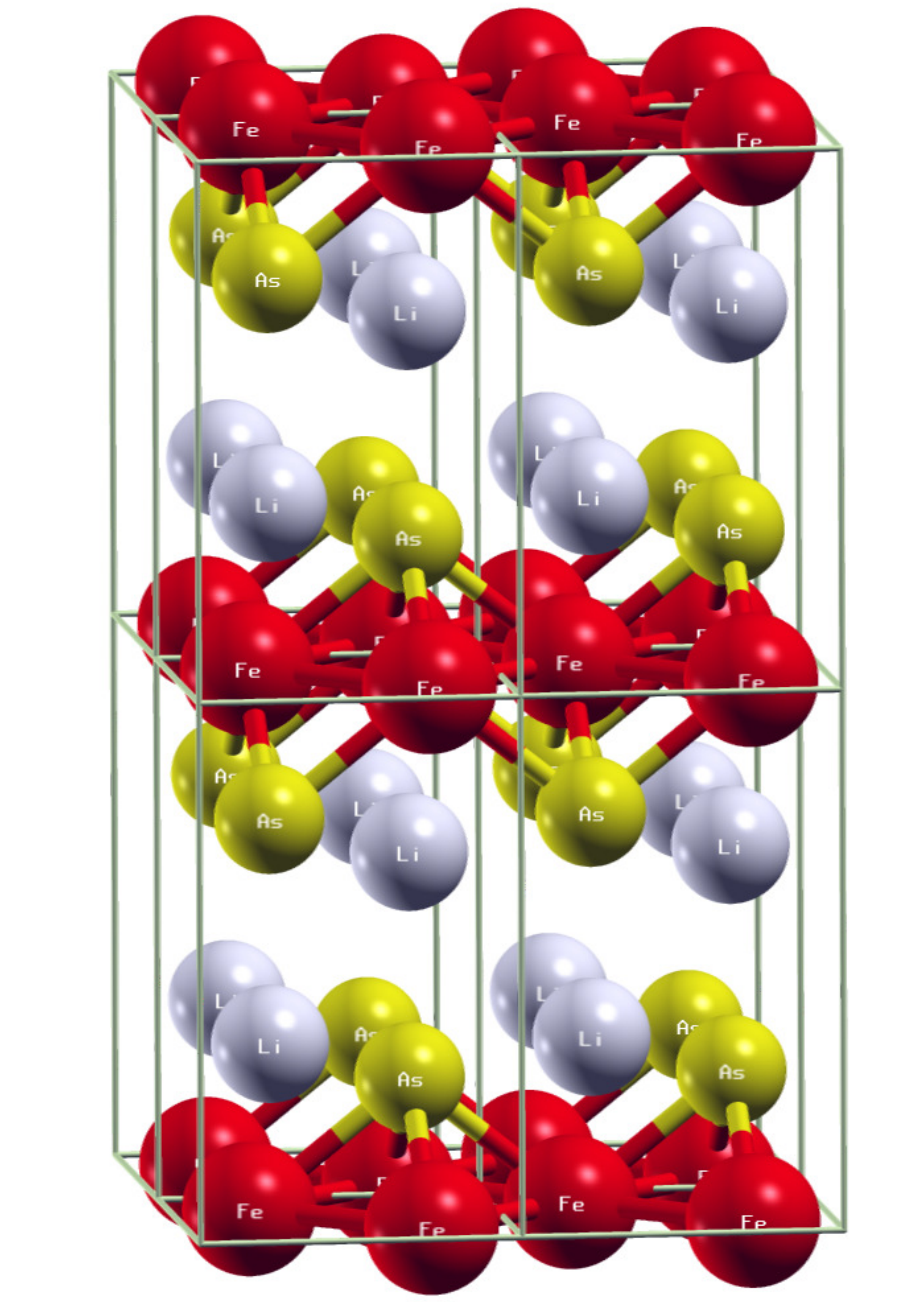} 
\end{center}
\caption{(Color online) Crystal structure of LiFeAs. The structure is tetragonal with space group $P4/nmm$. Iron atoms are represented by red spheres, arsenic  by yellow spheres. Light grey balls are larger cations, Li, between the layers. The $c$-axis points up. Note that a $2\times 2\times 2$ super-cell is plotted here. The FeAs layers are the electronically active part of the compound.}
\label{fig0-SM}
\end{figure}

LiFeAs crystallizes in a tetragonal structure with a space group $P4/nmm$. \fref{fig0-SM} shows the crystal structure of LiFeAs. In our study the crystal structure  is fixed to the experimental structure. 

We preformed a fully self-consistent LDA+DMFT calculation.~\cite{PhysRevB.81.195107-SM} We use a local Coulomb integral $U=F_0=4.0$~eV and a Hund's coupling $J=(F_2+F_4)/14=0.8$~eV, where $F_k$ are Slater integrals.~\cite{PhysRevB.85.094505-SM} We take the fully localized limit double-counting correction. We use $32\times 32\times 16$ ${\bf k}$-points and a rotationally invariant interaction matrix. 

\begin{figure}
\begin{center}
\begin{tabular}{c}
\includegraphics[width=0.69\columnwidth]{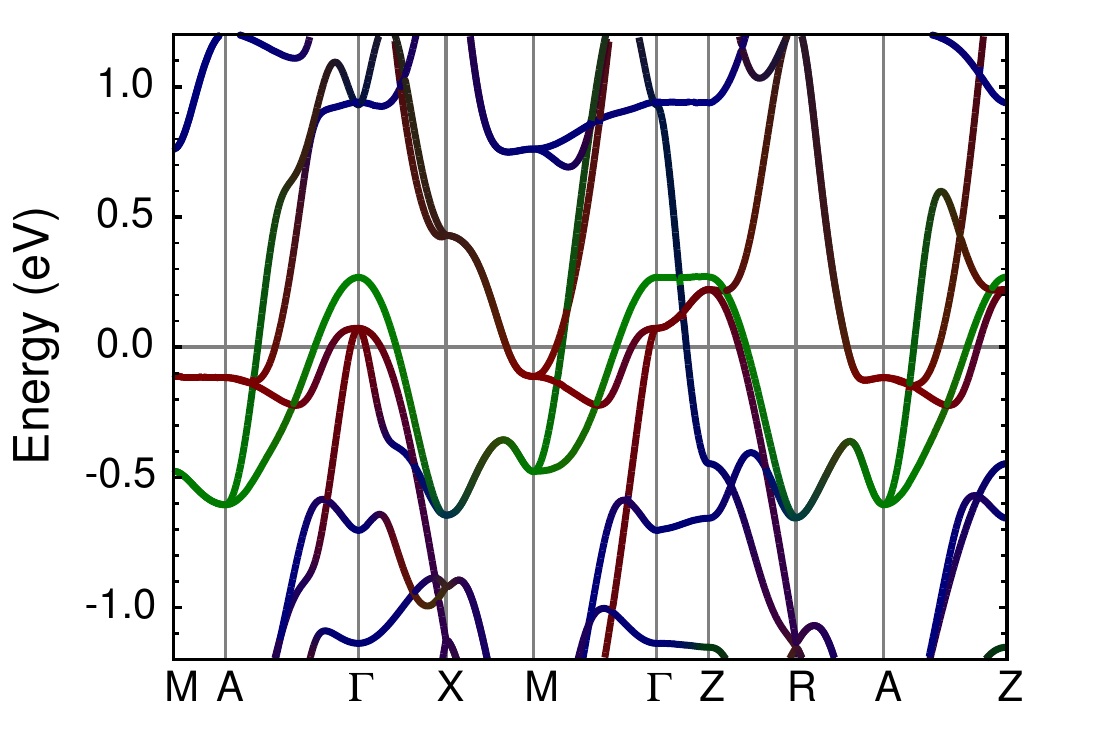} \\
\includegraphics[width=0.7\columnwidth]{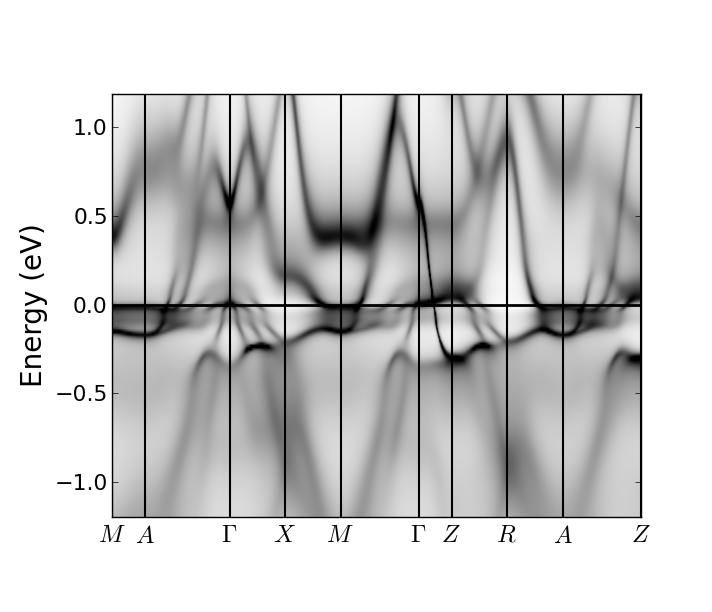} 
\end{tabular}
\end{center}
\caption{(Color online) LDA electronic band-structure (top) and LDA+DMFT momentum-resolved spectral function (bottom).~\cite{PhysRevB.81.195107-SM} The colors indicate Fe-$d$ and As-$p$ orbital contents: $d_{xz(yz)}$, $d_{xy}$, are denoted by red and green respectively, while all other orbitals are displayed in blue.}
\label{fig1-SM}
\end{figure}

The momentum-resolved spectral function, ${\rm Tr}{\bf A}({\bf k}, \omega)$, (trace over all orbitals) along high-symmetry directions is displayed in \fref{fig1-SM}, and compared with the LDA band structure. 
Compared to other pnictide families, LiFeAs has much shallower hole pockets around $\Gamma$. The flat top of these pockets implies a large density of states,~\cite{PhysRevLett.105.067002-SM} which has been proposed to promote ferromagnetic fluctuations and triplet $p$-wave pairing.~\cite{PhysRevB.83.060501-SM} However, experiment supports singlet-pairing. 

As can be seen from ${\rm Tr}{\bf A}({\bf k}, \omega)$, correlation effects lead to bandwidth reduction for the Fe-$3d$ states near the Fermi level: both the hole-pocket near the $\Gamma$ point and the electron-pocket near the $M$-point are pushed towards the Fermi energy. The outer hole pocket becomes more incoherent. 

The two low-energy hole-like bands at the $\Gamma$ point form a degenerate doublet with the two-dimensional irreducible representation $E_g$.~\cite{PhysRevB.88.134510-SM} The subscript $g$ and $u$ denote the spatial even and odd parity respectively. The two degenerate eigenstates are predominantly given by $( ic_1(d^A_{xz}+d^B_{xz})-c_2(d^A_{yz}+d^B_{yz}), ic_2(d^A_{xz}+d^B_{xz})+c_1(d^A_{yz}+d^B_{yz}))$ where  $A$, $B$ refer to Fe ions and $c_1$, $c_2$ are real coefficients. 
Away from the $\Gamma$ point, the symmetry reduces and the doublet splits. Depending on the $\bf k$-point, the eigenstate is composed predominantly from $d_{xz}$, or $d_{yz}$ orbital, but a phase factor $i$ between expansion coefficients is present everywhere. 
Along the high-symmetry lines $\Gamma$-$M$ and $\Gamma$-$X$, the $E_g$ symmetry reduces to one $A_2$ and one $B_2$ state. 


The outer hole-like band at the $\Gamma$-point instead has $B_{1g}$ symmetry with the $d^A_{xy}+d^B_{xy}$ Fe states. 
Along the high-symmetry lines $\Gamma$-$M$ and $\Gamma$-$X$, the symmetry reduces to $B_1$ and $A_1$, respectively.~\cite{PhysRevB.88.134510-SM} 
All the irreducible representations at the BZ boundary are two-dimensional~\cite{PhysRevB.88.134510-SM} which implies that all the Bloch states at any $\bf k$ on the boundary, including $M$-$X$ line, are doubly degenerate. This can be seen from \fref{fig1-SM}. At the $M$-point, the low-energy doubly degenerate electron-like bands are composed of ($d^A_{xz}+id^B_{yz}$,$d^A_{yz}-id^B_{xz}$)
with symmetry $A_2 \oplus B_2$ along $M$-$\Gamma$ and $E$ along $M$-$X$. 
The doubly degenerate bands with higher binding energy are composed of ($d^A_{xy}+d^B_{xy}$,$d^A_{xy}-d^B_{xy}$) with symmetry $B_1 \oplus B_2$ and $E$ along $M$-$\Gamma$ and  $M$-$X$, respectively.~\cite{PhysRevB.88.134510-SM}

\begin{figure}
\begin{center}
\begin{tabular}{ccc}
\includegraphics[width=0.32\columnwidth]{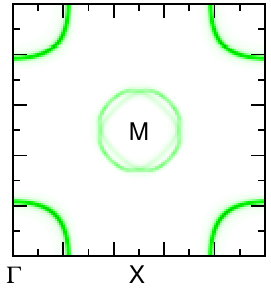} &
\includegraphics[width=0.32\columnwidth]{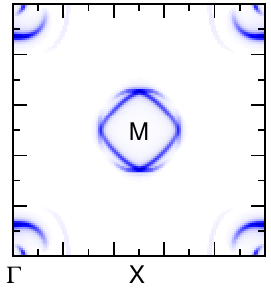} &
\includegraphics[width=0.32\columnwidth]{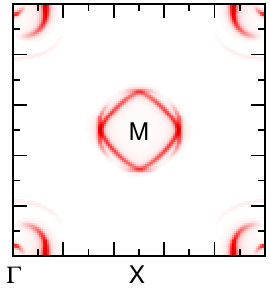} \\
\includegraphics[width=0.32\columnwidth]{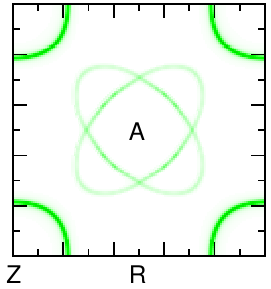} &
\includegraphics[width=0.32\columnwidth]{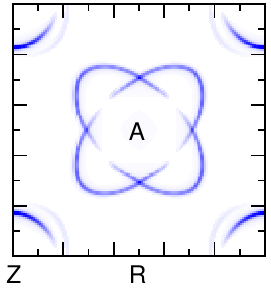} &
\includegraphics[width=0.32\columnwidth]{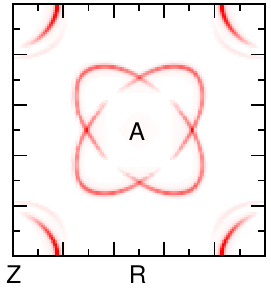}
\end{tabular}
\end{center}
\caption{(Color online) Fe $t_{2g}$ character of the FSs at $k_x$-$k_y$
plan with $k_z = 0$ (top panels), and $k_z = \pi/c$ (bottom panels) obtained from LDA calculation. Left, middle and right columns show $d_{xy}$, $d_{xz}$  and $d_{yz}$ orbitals respectively. }
\label{fig2-1-SM}
\end{figure}

\begin{figure}
\begin{center}
\begin{tabular}{ccc}
\includegraphics[width=0.32\columnwidth]{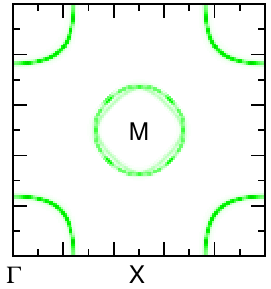} &
\includegraphics[width=0.32\columnwidth]{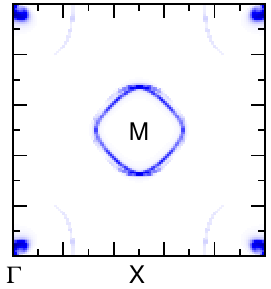} &
\includegraphics[width=0.32\columnwidth]{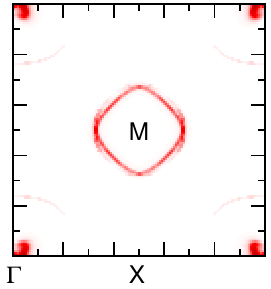} \\
\includegraphics[width=0.32\columnwidth]{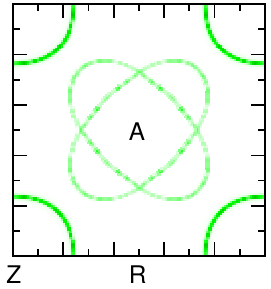} &
\includegraphics[width=0.32\columnwidth]{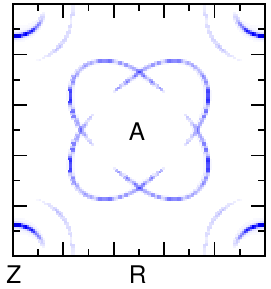} &
\includegraphics[width=0.32\columnwidth]{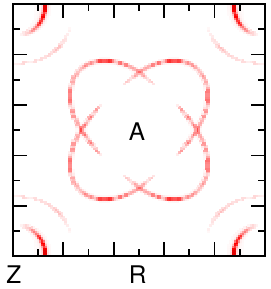}
\end{tabular}
\end{center}
\caption{(Color online) Fe $t_{2g}$ character of the FSs in the $k_x$-$k_y$
plane with $k_z = 0$ (top panels), and $k_z = \pi/c$ (bottom panels) obtained from the LDA+DMFT calculation. Left, middle and right columns show $d_{xy}$, $d_{xz}$  and $d_{yz}$ orbitals. The inner hole pocket crosses the Fermi level only in close vicinity to the $\Gamma$ point and is hard to resolve.}
\label{fig2-2-SM}
\end{figure}

\fref{fig2-1-SM} and \fref{fig2-2-SM} illustrate respectively the LDA and LDA+DMFT partial spectral weight, $A_{ll}({\bf k}, \omega)$, of Fe $t_{2g}$ orbitals on the Fermi surfaces. The two inner hole pockets are predominantly composed from $d_{xz}$ and $d_{yz}$ orbitals.  The smallest hole pocket 
has strong $k_z$ dispersion and is present only at small $k_z$, while the middle one has 
weak $k_z$ dispersion. The large hole-like Fermi surface originates purely from in-plane $d_{xy}$ orbitals and therefore is $2D$ without noticeable $k_z$ dispersion.  
The electron pockets are made from an admixture of  $d_{xy}$, $d_{xz}$ and $d_{yz}$ orbitals with weak $k_z$ dispersion. The electron pockets intersect at finite $k_z$ and their order flips, i.e., the inner pocket at $k_z=0$ is the outer pocket at $k_z=\pi/c$. The orbital character of portions of the $\beta_2$ pocket 
oriented towards $\Gamma$-points is $d_{xy}$, while for those portions oriented toward $X$-point it is $d_{xz}$ or $d_{yz}$. The $\beta_1$ pocket originates made predominantly from $d_{xz}$ and $d_{yz}$ orbitals.

Including interactions with LDA+DMFT, the two inner hole pockets shrink while the outer one expands. The middle hole pocket also deforms and takes on a butterfly shape at small $k_z$. The electron pockets are moderately expanded and they become closer to each other.


\section{Superconducting instability}
A superconducting instability in the singlet channel occurs when the pairing susceptibility in this channel diverges as one lowers temperature. A divergent susceptibility signals the appearance of a pole in the corresponding reducible complex vertex function, which  describes all scattering processes of two propagating particles. Using the Bethe-Salpeter equation, the reducible vertex function may be written as a function of the irreducible one and of the susceptibility that does not include vertex corrections. We will call this susceptibility bare, even though the Green functions can be dressed. One can then check that the condition for an instability is that an eignvalue of the matrix $-{\bm \Gamma}^{irr, s}{\bm \chi}^{0}_{pp}$ becomes unity. Here ${\bm \Gamma}^{irr, s}$ is the irreducible vertex function in the singlet channel, and ${\bm \chi}^{0}_{pp}$ is the bare susceptibility in the particle-particle channel.~\cite{Bickers2004-SM, PhysRevB.57.5376-SM} This yields a non-Hermitian eigenvalue problem, equivalent the linearized Eliashberg equation, defined as
  
%
\begin{align}
-&(\frac{k_BT}{N})^2\sum_{K'K'',l_3 \ldots l_6}
\left[{\bm \Gamma}^{irr, s}(Q=0)\right]_{K,l_1 l_2;K',l_3 l_4}
\times \nonumber\\
&\left[{\bm \chi}^{0}_{pp}(Q=0)\right]_{K',l_3 l_4;K'',l_5 l_6}
{\bm\Delta}_{K'',l_5l_6} =\lambda(T){\bm \Delta}_{K,l_1l_2},
\label{eq:EL1-SM}
\end{align}
where ${\bm \Delta}_{l_1,l_2}({\bf k},i\omega_m)$ is the gap function which evolves smoothly into the off-diagonal (anomalous) self-energy \emph{bellow} $T_c$.~\cite{PhysRevB.57.5376-SM} $l_1,\ldots,l_6$ are combined  ion (or sublattice) and orbital indices and we have defined $K \equiv ({\bf k},\omega_m)$ as momentum-frequency four-vectors. Even though the eigenvalue problem is non-Hermitian, the leading eigenvalues are real.~\cite{PhysRevB.57.5376-SM} The superconducting transition temperature, $T_c$, can be obtained as the temperature when the maximum (dimensionless) eigenvalue of \eref{eq:EL1-SM} becomes unity. The eigenvalue problem in \eref{eq:EL1-SM} involves diagonalization of a matrix of size $N\times (2N_{\omega}^{max}+2)\times N_{orb}^2$ where $N$ in the number of $\bf k$ points, $N_{\omega}^{max}$ in the maximum number of positive Matsubara frequencies and $N_{orb}$ denotes orbital numbers. 

The generalized bare susceptibility in the particle-particle channel is 
\begin{align}
\left[{\bm \chi}^{0}_{pp}(0)\right]_{K,l_1 l_2;K',l_3 l_4}=
\frac{N}{2k_BT}{\bf G}_{K,l_1l_3}{\bf G}_{-K,l_2l_4}\delta_{K,K'},
\label{eq:chipp-SM}
\end{align}
where ${\bf G}$ denotes the fully interacting propagator. A factor of $\frac{1}{2}$ in \eref{eq:chipp-SM} arises due to indistinguishability and should be considered to avoid double-counting.~\cite{Bickers2004-SM} This quantity is plotted in Fig.~(3) in the main text without the prefactor $N/k_BT$.

One way to see that the eigenvector of \eref{eq:EL1-SM} is related to the anomalous self-energy is as follows. Consider the one-band case in the BCS approximation. In this approximation one can neglect the frequency dependence of the gap function and of the pairing interaction, which allows to perform the summation on Matsubara frequencies.  Using bare propagators in \eref{eq:chipp-SM}, one can see that $[(k_BT)^2/N]\sum_{\omega_m}\left[{\bm \chi}^{0}_{pp}(0)\right]_{{\bf k},i\omega_m;-{\bf k},-i\omega_m}=(2n_F(\epsilon_{\bf k})-1)/4\epsilon_{\bf k}=\tanh(\epsilon_{\bf k}/2k_BT)/4\epsilon_{\bf k}$. Here, $n_F$ denotes the Fermi distribution function and we have used the relation $n_F(\epsilon_{\bf k})=(1-\tanh(\epsilon_{\bf k}/2k_BT))/2$. Inserting this into  \eref{eq:EL1-SM}, one finds
\begin{align}
-(\frac{1}{2N}) &\sum_{{\bf k}}
\left[{\bm \Gamma}^{irr, s}\right]_{{\bf k};{\bf k}'}
\frac{\tanh(\epsilon_{{\bf k}'}/2k_BT)}{2\epsilon_{{\bf k}'}}
{\bm\Delta}_{{\bf k}'} \nonumber \\&=\lambda(T){\bm \Delta}_{{\bf k}},
\label{eq:EL1BCS-SM}
\end{align}
which, for  $\lambda=1 $, is the \emph{linearlized} BCS gap equation.  The BCS gap equation has the same form as \eref{eq:EL1BCS-SM} but with $\epsilon_{\bf k}$ replaced by the quasi-particle excitation $E_{\bf k}=\sqrt{\epsilon^2_{\bf k}+|{\bm \Delta}_{{\bf k}}|^2}$.  Considering the Bogoliubov Hamiltonian, one can see that ${\bm \Delta}_{{\bf k}}$ is the anomalous self-energy.
 
\section{Pairing interaction}
Evaluation of the effective pairing interaction ${\bm \Gamma}^{irr,s}$ is a quantum many-body problem.~\cite{Bickers2004-SM, PhysRevB.57.5376-SM} 
One way is to employ Parquet equations to formally rewrite it in terms of the fully irreducible vertex function in the singlet channel, ${\bm \Lambda}^{irr,s}$, and of the vertex ladders in the particle-hole density and magnetic channels, ${\bm \Phi}^{d/m}_{ph}$, as~\cite{Bickers2004, 1367-2630-11-2-025016-SM}
%
\begin{widetext}
\begin{align}
[{\bm \Gamma}^{irr,s}(Q)]_{Kl_1l_2;K'l_3l_4}&=[{\bm \Lambda}^{irr,s}(Q)]_{Kl_1l_2;K'l_3l_4}
-\frac{1}{2}[{\bm \Phi}^d_{ph}(K'-K)]_{-K'l_2l_4;K+Ql_3l_1}
+\frac{3}{2}[{\bm \Phi}^m_{ph}(K'-K)]_{-K'l_2l_4;K+Ql_3l_1}\nonumber\\
&-\frac{1}{2}[{\bm \Phi}^d_{ph}(K'+K+Q)]_{-K'l_1l_4;-Kl_3l_2}
+\frac{3}{2}[{\bm \Phi}^m_{ph}(K'+K+Q)]_{-K'l_1l_4;-Kl_3l_2},\label{eq:Parquet-SM}
\end{align}
\end{widetext}
where a negative sign difference between terms including ${\bm \Phi}^{d/m}_{ph}$ and the corresponding terms in Ref.~\cite{Bickers2004-SM}  is compensated by a negative sign difference in definition of susceptibilities (see \eref{eq:dsus-SM} below).

The momentum-frequency appearing as an argument shows the transferred momentum-frequency, while the two momentum-frequency appearing as indices indicate the momentum-frequency of one of the incoming external legs and one of the outgoing ones. In other words, $[{\bm \Gamma}^{irr,s}(Q)]_{Kl_1l_2;K'l_3l_4}$ describes the particle-particle scattering of electrons in orbitals $l_1$, $l_2$ with momenta/frequencies ($K+Q$,$-K$) to electrons in orbitals $l_3$, $l_4$ with momenta/frequencies ($K'+Q$,$-K'$). Here, we are interested in the pairing of electrons at zero center of mass momentum and zero Matsubara frequency, i.e., $Q=0$ component. 

The vertex $[{\bm \Lambda}^{irr,s}(Q)]_{Kl_1l_2;K'l_3l_4}$ is irreducible in all two-particle channels. It is not singular. Its leading order is given by a linear combination of the irreducible density and magnetic vertices. The vertex ladder functions that appear in \eref{eq:Parquet-SM} are important for pairing and are defined as
%
%
%
\begin{align}
{\bm \Phi}^{d/m}_{ph}(Q)={\bm \Gamma}^{irr,d/m}(Q)
{\bm \chi}_{ph}^{d/m}(Q){\bm \Gamma}^{irr,d/m}(Q),\label{eq:dsus-SM}
\end{align}
where the right-hand side is a matrix multiplication in $K$ and orbital indices combined. Each matrix multiplication has an implied normalization factor $(k_BT/N)$. The functions ${\bm \Gamma}^{irr,d(m)}$ and ${\bm \chi}_{ph}^{d/m}$ are, respectively, irreducible vertex functions and generalized dressed susceptibilities in the density/magnetic channels. Using the Bethe-Salpeter equation in the particle-hole density/magnetic channels, ${\bm \chi}^{d/m}$ can be decomposed into bare susceptibility and vertex correction with the help of the irreducible vertex functions:
%
%
%
\begin{align}
{\bm \chi}_{ph}^{d(m)}(Q) &= {\bm \chi}^{0}_{ph}(Q)\nonumber\\
&-(+){\bm \chi}_{ph}^{d(m)}(Q)
{\bm \Gamma}^{irr,d(m)}(Q){\bm \chi}^{0}_{ph}(Q).
\label{eq:BSSus-SM}
\end{align}
Again the second term on the right-hand side is a matrix multiplication with the corresponding normalization factor and the bare generalized susceptibility is given by
\begin{equation}
[{\bm \chi}^{0}_{ph}(Q)]_{Kl_1l_2;K'l_3l_4} = - (\frac{N}{k_BT}) {\bf G}_{K+Q,l_1l_3}{\bf G}_{K,l_4l_2}\delta_{K,K'}.
\end{equation}

For further progress, one has to do some approximation. 
One can calculate the ladder functions in the particle-hole channel, ${\bm \Phi}^{d/m}_{ph}$, by simply replacing all irreducible vertex functions with a static and momentum-independent interaction. This corresponds to the random-phase approximation (RPA). 
In RPA another simplification appears due to time- and spatial-locality of the vertex functions. This allows to perform the summations over momentum-frequency of the external legs. This greatly simplifies the equations and it also reduces the size of matrices, which now are only functions of transferred momentum-frequency vector and of orbital indices. For example, for the bare susceptibility in the particle-hole channel we obtain
\begin{align}
[{\bm \chi}^{0}_{ph}(Q)]_{l_1l_2;l_3l_4} 
&= -(\frac{k_BT}{N})^2\sum_{KK'} [{\bm \chi}^{0}_{ph}(Q)]_{Kl_1l_2;K'l_3l_4} \nonumber\\
&=- (\frac{k_BT}{N})\sum_K {\bf G}_{K+Q,l_1l_3}{\bf G}_{K,l_4l_2}.\label{eq:bareSusRPA-SM}
\end{align}
Thus, one can drop the external momentum-frequency in the Parquet equations and rewrite the pairing interaction at $Q=0$ as
%
\begin{align}
[&{\bm \Gamma}^{irr,s}(0)]_{Kl_1l_2;K'l_3l_4}=[{\bm \Lambda}^{irr,s}(0)]_{Kl_1l_2;K'l_3l_4}\nonumber\\
&-\frac{1}{2}[{\bm \Phi}^d_{ph}(K'-K)]_{l_2l_4;l_3l_1}
+\frac{3}{2}[{\bm \Phi}^m_{ph}(K'-K)]_{l_2l_4;l_3l_1}\nonumber\\
&-\frac{1}{2}[{\bm \Phi}^d_{ph}(K'+K)]_{l_1l_4;l_3l_2}
+\frac{3}{2}[{\bm \Phi}^m_{ph}(K'+K)]_{l_1l_4;l_3l_2}.\label{eq:ParquetRPA-SM}
\end{align}
%


In the DMFT framework, the irreducible vertices are still local but they become frequency dependent~\cite{PhysRevB.75.045118-SM, PhysRevB.86.125114-SM} and in the context of iron pnictide superconductivity have only been calculated in the LDA+DMFT study of Refs.~\cite{Nat.Phys.10.845-SM, PhysRevB.86.064411-SM}. However, obtaining the full frequency dependence of the DMFT vertex is numerically very expensive and challenging for the multiorbital systems at the low temperatures. A calculation of the dressed susceptibility in DMFT is possible by quantum Monte Carlo (QMC) in the Hirsch-Fye algorithm~\cite{PhysRevLett.56.2521-SM} or hybridization expansion algorithm~\cite{RevModPhys.83.349-SM}.  The irreducible vertex can be obtained from the dressed susceptibility using Bethe-Salpeter equation. 
However, both methods have serious limitations. The Hirsch-Fye algorithm is only applicable at very high temperature, which is not suitable for studying superconductivity. In the temperature range accessible by this method, the pairing instability is very weak. The hybridization expansion algorithm gives good results for the dressed susceptibility only at low frequencies. Indeed, the noise grows very fast for moderate and high frequencies.~\cite{PhysRevB.85.205106-SM} To extract the irreducible vertex from the dressed susceptibility, using the Bethe-Salpeter equation, requires all frequencies, consequently, even the low frequency data of the full vertex are not usable.   
Furthermore, even if we assume that we have the frequency dependent irreducible vertex, the size of eigenvalue problem in the Eliashberg equation for multi-band systems is so large that only a limited number of lowest Matsubara frequencies can be considered. On this limited number of frequencies, the irreducible vertex does not reach its asymptotic behavior. Therefore an RPA analysis with static screened irreducible vertices (representing an average over frequency) seems a better approximation. Hence we take local frequency-independent irreducible vertices.~\cite{PhysRevB.75.134519-SM} Since the pnictides are weakly to moderately correlated systems, including the DMFT propagators in the resulting RPA-like equations gives a better description of
Fermi surface nesting, and accounts for the interaction of propagating particles with their environment.  We think the results should be more reliable than standard LDA+RPA.

Even with the above approximations for ${\bm \Phi}^{d/m}_{ph}$ the equation for ${\bm \Gamma}^{irr,s}$ depends on the incoming and outcoming momentum-frequency vectors. Thus, full matrices appear in the Eliashberg equation. 

To our knowledge, in all the previous studies another approximation, the so called BCS approximation, has been used to solve the Eliashberg equation. In this approximation, since the pairing interaction falls off rapidly on a frequency scale which is small compared with the bandwidth,  one can neglect the frequency dependence of the gap function. 
Furthermore, in this approximation one takes into account only the contribution from the lowest Matsubara frequency, the effective interaction is symmetrized, and its imaginary part is neglected,~\cite{PhysRevB.88.174516-SM} making the Eliashberg equation a real symmetric equation. Then the summation over Matsubara frequencies in the bare p-p susceptibility is performed to further simplify the Eliashberg equation. Moreover, often the resulting equation is projected on the Fermi surfaces and is solved assuming a real gap function with a specific symmetry.~\cite{1367-2630-11-2-025016-SM, PhysRevB.89.144513-SM} Strong coupling treatments such as those of Refs.~\cite{Nat.Commun.4.2783-SM} and  \cite{Sci.Rep.2.381-SM} do not restrict the pairing function to the vicinity of the Fermi surface but do not take into account the retardation of the pairing interaction.

Here, we solve the full equation in the physical $2$-Fe Brillouin zone with $10$-d orbitals at $k_BT=0.01$~eV. We use a $16\times 16\times 2$  ${\bf k}$-mesh and $12$ Matsubara frequencies. The convergence of the results have been checked by changing the above values to a $8\times 8\times 2$  ${\bf k}$-mesh and $24$ Matsubara frequencies. The three largest eigenvalues of the former set are $\simeq 0.125, 0.099, 0.085$ and they change by less than $10\%$ for the second set. More importantly, the gap symmetries do not change between the two sets.  The leading channel is conventional $s^{+-}$, the next-leading channel is anti-phase $s^{+-}$ while the third one has $d_{xy}$ symmetry.  

The irreducible vertex function in density/magnetic channels are defined as: ${\bm \Gamma}^{irr,d(m)}={\bm \Gamma}^{irr,\uparrow \downarrow}+(-){\bm \Gamma}^{irr,\uparrow \uparrow}$. In RPA, the irreducible vertex function is replaced by the antisymmetrized static Coulomb vertex, ${\bm \Gamma}^{0,\sigma\sigma'}$ which, in terms of the interacting part of Hamiltonian 
$(1/2)\sum_i\sum_{l_1 \ldots l_4}\sum_{\sigma\sigma'}I^{\sigma\sigma'}_{l_1l_2,l_3l_4}c^{\dagger}_{il_1\sigma}c^{\dagger}_{il_2\sigma'}
c_{il_3\sigma'}c_{il_4\sigma}$ is defined by ${\bm \Gamma}^{0,\sigma \sigma}_{l_1l_2;l_3l_4}=I^{\sigma\sigma}_{l_1l_4,l_3l_2}-I^{\sigma\sigma}_{l_1l_4,l_2l_3}$ and  ${\bm \Gamma}^{0,\sigma \bar{\sigma}}_{l_1l_2;l_3l_4}=I^{\sigma\bar{\sigma}}_{l_1l_4,l_3l_2}$, where $\bar{\sigma}\equiv -\sigma$. For a local interaction the following forms for density/magnetic  irreducible vertex functions are obtained
\begin{align}
{\bm \Gamma}^{irr,d(m)}_{l_1l_2;l_3l_4} = \left\{
  \begin{array}{l l l l l}
    U_s(U_s) & \quad l_1=l_2=l_3=l_4\\
    -U_s'+2J_s(U_s') & \quad l_1=l_3\neq l_2=l_4 \\
    2U_s'-J_s(J_s) & \quad l_1=l_2\neq l_3=l_4 \\
    J_s'(J_s')   & \quad l_1=l_4\neq l_2=l_3 \\
    0         & \quad \text{otherwise}
  \end{array} \right.
  \label{IntPam-SM}
\end{align} 
where $U_s$ and $U_s'$ denote the screened static local intra- and inter-orbital density-density interactions while $J_s$ and $J_s'$ are Hund's coupling and pair-hopping interactions. Due to locality of the interaction the four orbital indices belong to same ion. We also assume spin rotational invariance so the equalities $U_s'=U_s-\frac{5}{2}J_s$~\cite{0034-4885-74-12-124508-SM} and $J_s'=J_s$ are satisfied. Finally, the fully irreducible vertex $[{\bm \Lambda}^{irr,s}(Q)]_{Kl_1l_2;K'l_3l_4}$ is replaced by $\frac{1}{2}\left({\bm \Gamma}^{irr,d}+{\bm \Gamma}^{irr,m}\right)$ transformed to the particle-particle channel.

In our study we used two sets of screened interaction parameters yielding the same magnetic Stoner factor, namely $J_s=0.1U_s$, $U_s=2.4$~eV and $J_s=0.3U_s$, $U_s=1.68$~eV. 
The values for $U_s$ and $J_s$ are relatively  standard  in  the  literature that uses the RPA approach for the pairing vertex. They are smaller than cRPA values~\cite{doi:10.1143/JPSJ.79.044705-SM} in order to account for the RPA overestimation of spin fluctuations. We found that for a given $J_s/U_s$ ratio changing them within a limited range does not change the qualitative aspects of our results for the superconducting state. 

\section{RPA Susceptibilities}
\begin{figure}
\includegraphics[width=0.6\columnwidth]{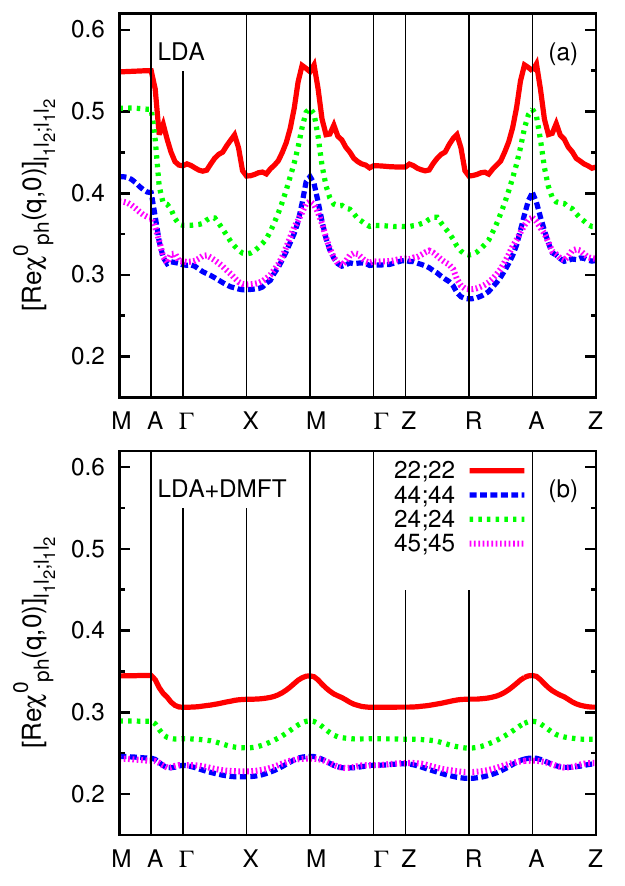}
\caption{(Color online) comparison between LDA (panel (a)) and DMFT (panel (b)) components of the bare susceptibility, $[{\bm \chi}^{0}_{ph}({\bf q},\nu_n=0)]_{l_1l_2;l_1l_2}$, of LiFeAs at $k_BT=0.01$~eV in the particle-hole channel (top panels). For $K$ summation in \eref{eq:bareSusRPA-SM} we have used a $32\times 32 \times 16$ $\bf k$-mesh and $1024$ positive frequencies. 
 }\label{fig3-SM}
\end{figure}

The top panels of \fref{fig3-SM} compare important components of the LDA bare susceptibilities in the particle-hole channel, ${\bm \chi}^{0}_{ph}({\bf q},\nu_n=0)$ at $k_BT=0.01$~eV, with their LDA+DMFT counterpart along a high-symmetry path. The Fe-1 (Fe-2) $d_{z^2}$, $d_{xy}$, $d_{x^2-y^2}$, $d_{xz}$, $d_{yz}$ 
orbitals are labeled respectively as $1$~($6$), $2$~($7$), $3$~($8$), $4$~($9$), and $5$~($10$). Due to better nesting,~\cite{PhysRevB.85.094509-SM} the dominant component of the LDA bare susceptibility is the $d_{xy}$ intra-orbital component, $[{\bm \chi}^{0}_{ph}]_{22;22}$ ($=[{\bm \chi}^{0}_{ph}]_{77;77}=[{\bm \chi}^{0}_{ph}]_{27;27}$), with incommensurate peaks around the $M$ ($A$) point, which come from nesting between the hole and electron pockets. The inter-orbital $d_{xy}$-$d_{xz(yz)}$ component, $(24;24)$ and $(25;25)$ respectively, is rather large and shows a peak centered at the $M$ ($A$) point. The intra-orbital $d_{xz(yz)}$ components $(44;44)$, and the inter-orbital  $d_{xz}$-$d_{yz}$ components $(45;45)$, have almost the same magnitude in the entire ${\bf k}$-space. Nesting between the hole (electron) pockets is weak and gives small peaks at small momenta.   

Upon introducing correlations within the LDA+DMFT framework, the absolute values of the bare susceptibilities are suppressed for all components. However, the AF fluctuations remain the leading instability. The small peaks due to intra-hole (electron) pockets scattering are shifted towards smaller momenta and smoothed, resulting in a relative increase of zero momentum scattering. 

\begin{figure}
\includegraphics[width=\columnwidth]{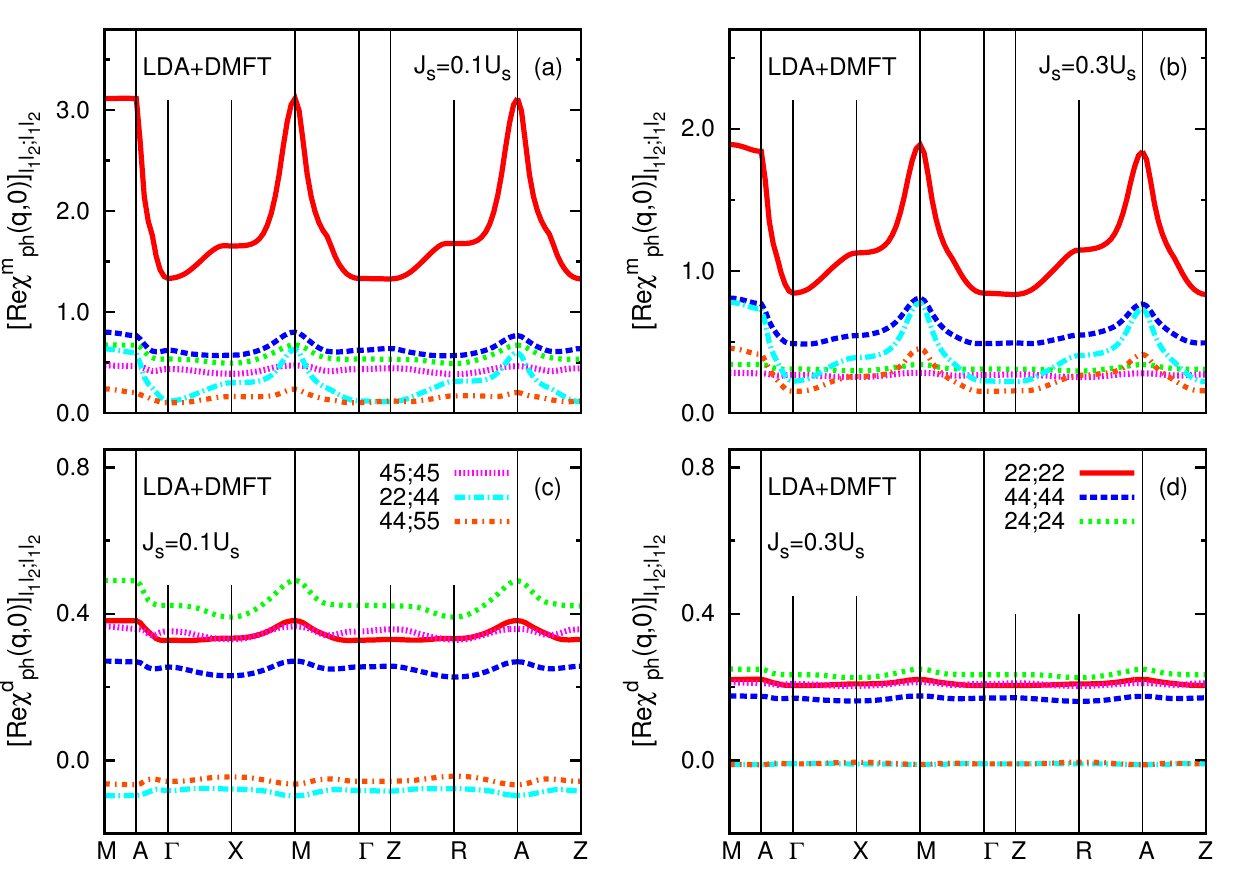}
\caption{(Color online) Several components of the dressed susceptibility in, respectively, the magnetic channel, density channel of LiFeAs at $k_BT=0.01$~eV in the particle-hole channel. There are two sets of screened interaction parameters yielding the same magnetic Stoner factor, namely $J_s=0.1U_s$, $U_s=2.4$~eV on the left and $J_s=0.3U_s$, $U_s=1.68$~eV on the right. Here, $J_s$ denotes Hund's coupling and $U_s$ the local intra-orbital Hubbard interaction. The inter-orbital interaction and pair hopping are determined assuming spin-rotational symmetry.  The same line color convention is used in all panels. The convention can be read from the insets of parts (c),(d) of the figure. 2,4,5 stand respectively for $d_{xy}$,$d_{xz}$ and $d_{yz}$.
 }\label{fig4-SM}
\end{figure}

\fref{fig4-SM} show several components of the RPA susceptibility in the density and magnetic channels for two sets of screened interaction parameters that yield the same magnetic Stoner factor. The RPA magnetic (charge)-susceptibility is similar to the bare one but enhanced (reduced) due to the effect of the Coulomb interaction, ${\bm\chi}_{ph}^{m(d)}={\bm\chi}_{ph}^0/[{\bm 1}-(+){\bm\chi}_{ph}^0{\bm\Gamma}^{m(d)}]$. The form of the irreducible vertex functions, \eref{IntPam-SM}, shows that a non-zero Hund's coupling breaks spin-orbital symmetry and suppresses orbital fluctuations. This can be seen from a suppression of the charge fluctuation upon increasing $J_s$ ( compare panels (c) and (d) of the \fref{fig4-SM}). The magnetic susceptibility is also clearly larger than the density susceptibility. This translates into a repulsive effective pairing interaction. This is more pronounced for the intra-orbital components in general and in particular for $d_{xy}$. 
The inter-orbital components $[{\bm\chi}^{m(d)}_{ph}]_{l_1l_2;l_1l_2}$, with for example $l_1,l_2$ given by $(24)$ $d_{xy}$-$d_{xz}$ or $(45)$ $d_{xz}$-$d_{yz}$  in \fref{fig4-SM}, are susceptibilities which in RPA are enhanced by inter-orbital Coulomb interaction, as can be seen by comparing with \fref{fig3-SM}(b).

Interestingly, upon increasing the $J_s/U_s$ ratio from \fref{fig4-SM}(a) to \fref{fig4-SM}(b), the $d_{xy}$ intra-orbital component $[{\bm\chi}^{m}_{ph}]_{22;22}$  decreases while the $d_{xy}$-$d_{xz(yz)}$ and $d_{xz}$-$d_{yz}$ inter-orbital components  $[{\bm\chi}^{m}_{ph}]_{22;44(55)}$ and $[{\bm\chi}^{m}_{ph}]_{44;55}$  increase slightly. Recall that the calculation is done for a given Stoner factor (proximity to magnetic transition). For $J_s/U_s=0.3$ the value of $U_s$ is smaller than for $J_s/U_s=0.1$. One would have expected that with a smaller $U_s$ intra-orbital magnetic correlations would have decreased for all orbitals. Instead, only $d_{xy}$ seems strongly affected. This suggests that the larger value of $J_s$ not only increases inter-orbital spin susceptibility, as expected, it also increases intra-orbital spin susceptibility in the less strongly correlated orbitals. It is as if, through $J_s$, spin fluctuations in the more correlated $d_{xy}$ orbital increase those in the other two $t_{2g}$ orbitals. This behavior of the magnetic susceptibility reflects itself directly in the pairing interaction. Furthermore, due to the negative sign of the corresponding off-diagonal dressed susceptibility in the charge channel, the magnetic and charge channels for the inter-orbital components cooperate to boost the corresponding pairing vertex.

\begin{figure}
\includegraphics[width=\linewidth]{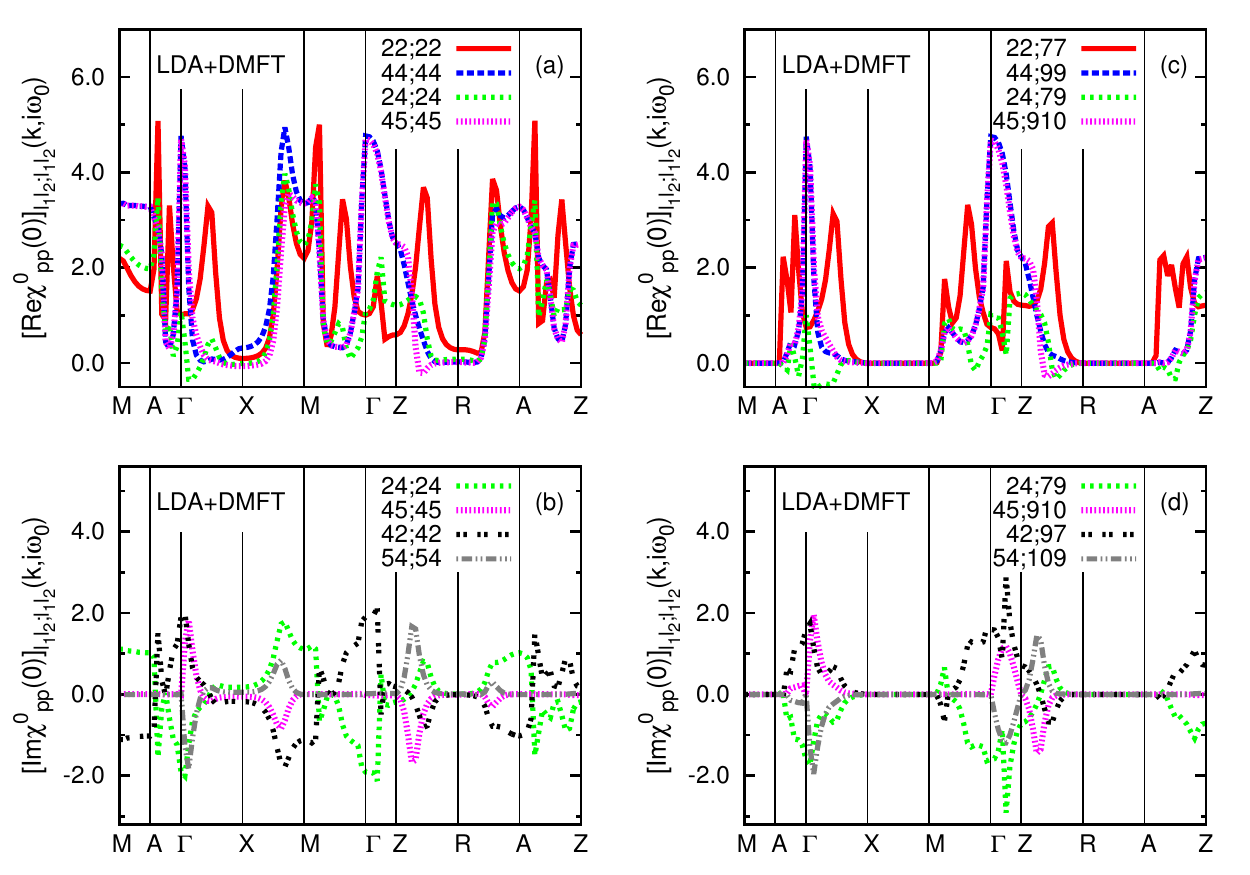}
\caption{(Color online) Real part (top) and imaginary part (bottom) of the several intra-sublattice (left) and inter-sublattice (right) components of the generalized particle-particle bare susceptibility at the lowest fermionic/bosonic Matsubara frequency. 
 }\label{fig5-SM}
\end{figure}

\paragraph{Bare particle-particle susceptibility}
The generalized bare susceptibility in the p-p channel also enters the gap equation.~\fref{fig5-SM} shows the real part in (a, c) and the imaginary part in (b, d) for several components of the generalized p-p bare susceptibility at the lowest fermionic/bosonic frequencies. The intra-orbital components are purely real. Both real and imaginary parts show relatively sharp peaks at the position of FSs. Panels (c) and (d) show the inter-sublattice components. In the BCS approximation, only real parts survive for the components considered here, due to a summation over Matsubara frequencies. In this case, the inter-orbital pairing is suppressed. Including the imaginary part in the full gap equation changes this trend. The imaginary parts of the inter-orbital components change sign between corner and center of the BZ.  
They have some symmetries that transfer to the gap function: (i) They are odd under exchange of orbital indices (see \eref{id2-SM}), as can be seen comparing the green and black lines or the grey and purple lines in \fref{fig5-SM}(b); (ii) There is also a $\pi$ phase difference between the two Fe ions (not shown). 

\section{Inter-orbital pairing components and their effects on the SC gap magnitude }
\begin{figure}
	\begin{center}
			\includegraphics[width=0.95\linewidth]{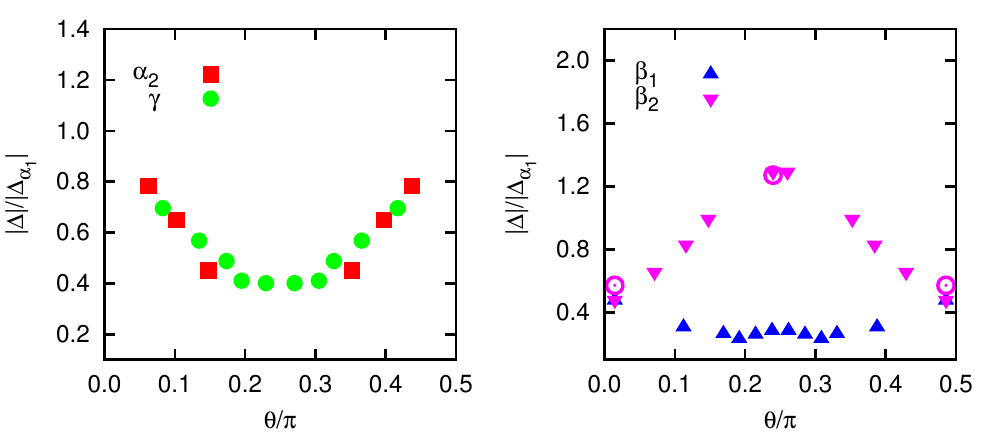} 
	\end{center}
	\caption{(Color online) For $J_s/U_s=0.3$, the SC gap magnitude (in units of the average gap magnitude on the $\alpha{_1}$ pocket) as a function of the angle $\theta$  measured at the $\Gamma$ and $M$ points with respect to the $k_x$ axis for $k_z=0$ FSs. Down filled pink triangles are for the $\beta_2$ pocket and up filled blue triangles are for the $\beta_1$ pocket. The open symbols show the gap magnitude at $\theta/\pi\simeq 0.0, 0.25, 0.5,$, calculated by setting the inter-orbital components to zero. This calculation qualitatively shows that the inter-orbital components reduce  the gap magnitude where the electron pockets intersect.}\label{fig6-SM}
\end{figure}

A quantitative evaluation of the inter-orbital pairing component effects on the SC gap magnitude requires a full disentanglement of the changes caused by the inter-orbital components. This is not a straightforward task, because inter-orbital components are coupled with intra-orbital components in the Eliashberg equations. One way to qualitatively understand their effects is to remove this component from the gap function and recalculate the gap magnitude. As can be seen from \fref{fig6-SM}, such an analysis shows that the inter-orbital components reduce the gap magnitude, in particular where the electron pockets intersect, which is where these components are maximum. It is worth mentioning that due to anisotropic spin fluctuations in our gauge,~\cite{1601.05813-SM} the gap function shows a very small anisotropy. The presented data here are symmetrized.

\section{Symmetries of Green's functions and Susceptibilities}
The normal Green's function ${\bf G}^{\sigma}_{l_1, l_2}({\bf k},\tau)$ describing the propagation of electrons from $l_1$ to $l_2$ with momentum $\bf k$ is defined as
\begin{align}
{\bf G}^{\sigma}_{l_1,  l_2}({\bf k},\tau)&=-\langle T_{\tau}c_{{\bf k} l_1\sigma}(\tau)c^{\dagger}_{{\bf k} l_2\sigma}(0)\rangle, \nonumber 
\end{align}
where $c_{{\bf k} l_1\sigma}(\tau)=\exp[(H-\mu N)\tau]c_{{\bf k}\alpha l_1\sigma}\exp[-(H-\mu N)\tau]$ with $\tau$ the imaginary time, $N=\sum_{{\bf k}\sigma}\sum_{l_1}c^{\dagger}_{{\bf k}l_1\sigma}c_{{\bf k}l_1\sigma}$ the number operator, $\mu$ the chemical potential, and $T_{\tau}$ the time-ordering operator. Define $K=H-\mu N$, and $\Omega$ as the grand potential. Then complex conjugation yields
\begin{align}
{\bf G}^{\sigma}_{l_1l_2}&({\bf k},\tau)^*=\nonumber \\
&\bigg(-\Theta(\tau){\rm Tr} \big[e^{-\beta(K-\Omega)}e^{K\tau} c_{{\bf k} l_1\sigma}e^{-K\tau}c^{\dagger}_{{\bf k} l_2\sigma}\big]^{\dagger}\nonumber \\
&+\Theta(-\tau){\rm Tr}\big[e^{-\beta(K-\Omega)}c^{\dagger}_{{\bf k} l_2\sigma} e^{K\tau} c_{{\bf k} l_1\sigma}e^{-K\tau}\big]^{\dagger} \bigg)\nonumber \\
&=-\langle T_{\tau}c_{{\bf k} l_2\sigma}(\tau)
c^{\dagger}_{{\bf k} l_1\sigma}(0)\rangle = {\bf G}^{\sigma}_{l_2l_1}({\bf k},\tau).
\end{align}
%

Transforming to Mastubara frequency space we obtain
\begin{align}
{\bf G}^{\sigma}_{l_1l_2}({\bf k}, i\omega_m)^*&={\bf G}^{\sigma}_{l_2l_1}({\bf k},-i\omega_m). 
\end{align}
This property of the Green's function implies for the bare susceptibility in the p-h channel, 
\begin{equation}
[{\bm \chi}^{0}_{ph}({\bf q},i\nu_n)]_{l_1l_2;l_3l_4}^* = [{\bm \chi}^{0}_{ph}({\bf q},-i\nu_n)]_{l_3l_4;l_1l_2}.
\end{equation}
It also gives a similar relation for the generalized bare susceptibility in the p-p channel with zero transferred momentum/frequency, $\left[{\bm \chi}^{0}_{pp}(0)\right]_{K,l_1 l_2;K',l_3 l_4}\equiv \left[{\bm \chi}^{0}_{pp}(0)\right]_{l_1 l_2;l_3 l_4}({\bf k},i\omega_m)$, as
\begin{equation}
\left[{\bm \chi}^{0}_{pp}(0)\right]_{l_1 l_2;l_3 l_4}({\bf k},i\omega_m)^* = \left[{\bm \chi}^{0}_{pp}(0)\right]_{l_3 l_4;l_1 l_2}({\bf k},-i\omega_m).
\end{equation}

If the time-reversal symmetry is respected then the Green's function transforms as 
\begin{align}
\mathcal{T}{\bf G}^{\sigma}_{l_1l_2}&({\bf k},\tau)=\nonumber\\ &-\Theta(\tau)e^{\beta \Omega}{\rm Tr} \big[c_{-{\bf k} l_2\bar{\sigma}}e^{-K\tau}
c^{\dagger}_{-{\bf k} l_1\bar{\sigma}}e^{K\tau}e^{-\beta K} 
\big]\nonumber\\
&+\Theta(-\tau)e^{\beta \Omega}{\rm Tr}\big[e^{-K\tau}c^{\dagger}_{-{\bf k} l_1\bar{\sigma}}e^{K\tau}
c_{-{\bf k} l_2\bar{\sigma}}e^{-\beta K} 
\big]\nonumber \\
&=-\langle T_{\tau}c_{-{\bf k} l_2\bar{\sigma}}(\tau)
c^{\dagger}_{-{\bf k} l_1\bar{\sigma}}(0)\rangle = {\bf G}^{\bar{\sigma}}_{ l_2 l_1}(-{\bf k},\tau)
\end{align}
where we have used
$\mathcal{T}|\uparrow\rangle = -ie^{-i\delta}|\downarrow\rangle$ and $\mathcal{T}|\downarrow\rangle = ie^{-i\delta}|\uparrow\rangle$ 
with $e^{-i\delta}=i$. Note that $d$-orbitals in cubic symmetry are time-reversal invariant. It can be seen from their definition in terms of $Y_l^m$.

Transforming to Mastubara frequency, the above equality is expressed in frequency space
as
\begin{align}
{\bf G}^{\sigma}_{l_1, l_2}({\bf k},i\omega_m)&={\bf G}^{\bar{\sigma}}_{l_2, l_1}(-{\bf k},i\omega_m). 
\end{align}

Combining with the earlier results obtained for complex conjugation, we obtain the following identity 
\begin{equation}
{\bf G}^{\sigma}_{l_1,l_2}({\bf k},i\omega_m)={\bf G}^{\bar{\sigma}}_{ l_1, l_2}(-{\bf k},-i\omega_m)^*,
\end{equation}
which for the generalized bare susceptibility in the p-p channel, \eref{eq:chipp-SM}, gives (assuming SU(2) symmetry)
\begin{align}
&\left[{\bm \chi}^{0}_{pp}(0)\right]_{K,l_1 l_2;K',l_1 l_2}=
\frac{N}{2k_BT}{\bf G}_{K,l_1l_1}{\bf G}_{-K,l_2l_2}\delta_{K,K'}\nonumber\\
&=\frac{N}{2k_BT}{\bf G}^*_{-K,l_1l_1}{\bf G}^*_{K,l_2l_2}\delta_{K,K'}=\left[{\bm \chi}^{0}_{pp}(0)\right]^*_{K,l_2 l_1;K',l_2 l_1}.\label{id2-SM}
\end{align}


\begin{thebibliography}{43}%
\makeatletter
\providecommand \@ifxundefined [1]{%
 \@ifx{#1\undefined}
}%
\providecommand \@ifnum [1]{%
 \ifnum #1\expandafter \@firstoftwo
 \else \expandafter \@secondoftwo
 \fi
}%
\providecommand \@ifx [1]{%
 \ifx #1\expandafter \@firstoftwo
 \else \expandafter \@secondoftwo
 \fi
}%
\providecommand \natexlab [1]{#1}%
\providecommand \enquote  [1]{``#1''}%
\providecommand \bibnamefont  [1]{#1}%
\providecommand \bibfnamefont [1]{#1}%
\providecommand \citenamefont [1]{#1}%
\providecommand \href@noop [0]{\@secondoftwo}%
\providecommand \href [0]{\begingroup \@sanitize@url \@href}%
\providecommand \@href[1]{\@@startlink{#1}\@@href}%
\providecommand \@@href[1]{\endgroup#1\@@endlink}%
\providecommand \@sanitize@url [0]{\catcode `\\12\catcode `\$12\catcode
  `\&12\catcode `\#12\catcode `\^12\catcode `\_12\catcode `\%12\relax}%
\providecommand \@@startlink[1]{}%
\providecommand \@@endlink[0]{}%
\providecommand \url  [0]{\begingroup\@sanitize@url \@url }%
\providecommand \@url [1]{\endgroup\@href {#1}{\urlprefix }}%
\providecommand \urlprefix  [0]{URL }%
\providecommand \Eprint [0]{\href }%
\providecommand \doibase [0]{http://dx.doi.org/}%
\providecommand \selectlanguage [0]{\@gobble}%
\providecommand \bibinfo  [0]{\@secondoftwo}%
\providecommand \bibfield  [0]{\@secondoftwo}%
\providecommand \translation [1]{[#1]}%
\providecommand \BibitemOpen [0]{}%
\providecommand \bibitemStop [0]{}%
\providecommand \bibitemNoStop [0]{.\EOS\space}%
\providecommand \EOS [0]{\spacefactor3000\relax}%
\providecommand \BibitemShut  [1]{\csname bibitem#1\endcsname}%
\let\auto@bib@innerbib\@empty
\bibitem [{\citenamefont {Tapp}\ \emph {et~al.}(2008)\citenamefont {Tapp},
  \citenamefont {Tang}, \citenamefont {Lv}, \citenamefont {Sasmal},
  \citenamefont {Lorenz}, \citenamefont {Chu},\ and\ \citenamefont
  {Guloy}}]{PhysRevB.78.060505}%
  \BibitemOpen
  \bibfield  {author} {\bibinfo {author} {\bibfnamefont {J.~H.}\ \bibnamefont
  {Tapp}}, \bibinfo {author} {\bibfnamefont {Z.}~\bibnamefont {Tang}}, \bibinfo
  {author} {\bibfnamefont {B.}~\bibnamefont {Lv}}, \bibinfo {author}
  {\bibfnamefont {K.}~\bibnamefont {Sasmal}}, \bibinfo {author} {\bibfnamefont
  {B.}~\bibnamefont {Lorenz}}, \bibinfo {author} {\bibfnamefont {P.~C.~W.}\
  \bibnamefont {Chu}}, \ and\ \bibinfo {author} {\bibfnamefont {A.~M.}\
  \bibnamefont {Guloy}},\ }\href {\doibase 10.1103/PhysRevB.78.060505}
  {\bibfield  {journal} {\bibinfo  {journal} {Phys. Rev. B}\ }\textbf {\bibinfo
  {volume} {78}},\ \bibinfo {pages} {060505} (\bibinfo {year}
  {2008})}\BibitemShut {NoStop}%
\bibitem [{\citenamefont {Allan}\ \emph {et~al.}(2012)\citenamefont {Allan},
  \citenamefont {Rost}, \citenamefont {Mackenzie}, \citenamefont {Xie},
  \citenamefont {Davis}, \citenamefont {Kihou}, \citenamefont {Lee},
  \citenamefont {Iyo}, \citenamefont {Eisaki},\ and\ \citenamefont
  {Chuang}}]{Allan04052012}%
  \BibitemOpen
  \bibfield  {author} {\bibinfo {author} {\bibfnamefont {M.~P.}\ \bibnamefont
  {Allan}}, \bibinfo {author} {\bibfnamefont {A.~W.}\ \bibnamefont {Rost}},
  \bibinfo {author} {\bibfnamefont {A.~P.}\ \bibnamefont {Mackenzie}}, \bibinfo
  {author} {\bibfnamefont {Y.}~\bibnamefont {Xie}}, \bibinfo {author}
  {\bibfnamefont {J.~C.}\ \bibnamefont {Davis}}, \bibinfo {author}
  {\bibfnamefont {K.}~\bibnamefont {Kihou}}, \bibinfo {author} {\bibfnamefont
  {C.~H.}\ \bibnamefont {Lee}}, \bibinfo {author} {\bibfnamefont
  {A.}~\bibnamefont {Iyo}}, \bibinfo {author} {\bibfnamefont {H.}~\bibnamefont
  {Eisaki}}, \ and\ \bibinfo {author} {\bibfnamefont {T.-M.}\ \bibnamefont
  {Chuang}},\ }\href {\doibase 10.1126/science.1218726} {\bibfield  {journal}
  {\bibinfo  {journal} {Science}\ }\textbf {\bibinfo {volume} {336}},\ \bibinfo
  {pages} {563} (\bibinfo {year} {2012})}
   \BibitemShut
  {NoStop}%
\bibitem [{\citenamefont {Borisenko}\ \emph {et~al.}(2010)\citenamefont
  {Borisenko}, \citenamefont {Zabolotnyy}, \citenamefont {Evtushinsky},
  \citenamefont {Kim}, \citenamefont {Morozov}, \citenamefont {Yaresko},
  \citenamefont {Kordyuk}, \citenamefont {Behr}, \citenamefont {Vasiliev},
  \citenamefont {Follath},\ and\ \citenamefont
  {B\"uchner}}]{PhysRevLett.105.067002}%
  \BibitemOpen
  \bibfield  {author} {\bibinfo {author} {\bibfnamefont {S.~V.}\ \bibnamefont
  {Borisenko}}, \bibinfo {author} {\bibfnamefont {V.~B.}\ \bibnamefont
  {Zabolotnyy}}, \bibinfo {author} {\bibfnamefont {D.~V.}\ \bibnamefont
  {Evtushinsky}}, \bibinfo {author} {\bibfnamefont {T.~K.}\ \bibnamefont
  {Kim}}, \bibinfo {author} {\bibfnamefont {I.~V.}\ \bibnamefont {Morozov}},
  \bibinfo {author} {\bibfnamefont {A.~N.}\ \bibnamefont {Yaresko}}, \bibinfo
  {author} {\bibfnamefont {A.~A.}\ \bibnamefont {Kordyuk}}, \bibinfo {author}
  {\bibfnamefont {G.}~\bibnamefont {Behr}}, \bibinfo {author} {\bibfnamefont
  {A.}~\bibnamefont {Vasiliev}}, \bibinfo {author} {\bibfnamefont
  {R.}~\bibnamefont {Follath}}, \ and\ \bibinfo {author} {\bibfnamefont
  {B.}~\bibnamefont {B\"uchner}},\ }\href {\doibase
  10.1103/PhysRevLett.105.067002} {\bibfield  {journal} {\bibinfo  {journal}
  {Phys. Rev. Lett.}\ }\textbf {\bibinfo {volume} {105}},\ \bibinfo {pages}
  {067002} (\bibinfo {year} {2010})}\BibitemShut {NoStop}%
\bibitem [{\citenamefont {Lee}\ \emph {et~al.}(2012)\citenamefont {Lee},
  \citenamefont {Ji}, \citenamefont {Kim}, \citenamefont {Kim}, \citenamefont
  {Haule}, \citenamefont {Kotliar}, \citenamefont {Lee}, \citenamefont {Khim},
  \citenamefont {Kim}, \citenamefont {Kim}, \citenamefont {Kim},\ and\
  \citenamefont {Shim}}]{PhysRevLett.109.177001}%
  \BibitemOpen
  \bibfield  {author} {\bibinfo {author} {\bibfnamefont {G.}~\bibnamefont
  {Lee}}, \bibinfo {author} {\bibfnamefont {H.~S.}\ \bibnamefont {Ji}},
  \bibinfo {author} {\bibfnamefont {Y.}~\bibnamefont {Kim}}, \bibinfo {author}
  {\bibfnamefont {C.}~\bibnamefont {Kim}}, \bibinfo {author} {\bibfnamefont
  {K.}~\bibnamefont {Haule}}, \bibinfo {author} {\bibfnamefont
  {G.}~\bibnamefont {Kotliar}}, \bibinfo {author} {\bibfnamefont
  {B.}~\bibnamefont {Lee}}, \bibinfo {author} {\bibfnamefont {S.}~\bibnamefont
  {Khim}}, \bibinfo {author} {\bibfnamefont {K.~H.}\ \bibnamefont {Kim}},
  \bibinfo {author} {\bibfnamefont {K.~S.}\ \bibnamefont {Kim}}, \bibinfo
  {author} {\bibfnamefont {K.-S.}\ \bibnamefont {Kim}}, \ and\ \bibinfo
  {author} {\bibfnamefont {J.~H.}\ \bibnamefont {Shim}},\ }\href {\doibase
  10.1103/PhysRevLett.109.177001} {\bibfield  {journal} {\bibinfo  {journal}
  {Phys. Rev. Lett.}\ }\textbf {\bibinfo {volume} {109}},\ \bibinfo {pages}
  {177001} (\bibinfo {year} {2012})}\BibitemShut {NoStop}%
\bibitem [{\citenamefont {Brydon}\ \emph {et~al.}(2011)\citenamefont {Brydon},
  \citenamefont {Daghofer}, \citenamefont {Timm},\ and\ \citenamefont {van~den
  Brink}}]{PhysRevB.83.060501}%
  \BibitemOpen
  \bibfield  {author} {\bibinfo {author} {\bibfnamefont {P.~M.~R.}\
  \bibnamefont {Brydon}}, \bibinfo {author} {\bibfnamefont {M.}~\bibnamefont
  {Daghofer}}, \bibinfo {author} {\bibfnamefont {C.}~\bibnamefont {Timm}}, \
  and\ \bibinfo {author} {\bibfnamefont {J.}~\bibnamefont {van~den Brink}},\
  }\href {\doibase 10.1103/PhysRevB.83.060501} {\bibfield  {journal} {\bibinfo
  {journal} {Phys. Rev. B}\ }\textbf {\bibinfo {volume} {83}},\ \bibinfo
  {pages} {060501} (\bibinfo {year} {2011})}\BibitemShut {NoStop}%
\bibitem [{\citenamefont {Allan}\ \emph {et~al.}(2015)\citenamefont {Allan},
  \citenamefont {Lee}, \citenamefont {Rost}, \citenamefont {Fischer},
  \citenamefont {Massee}, \citenamefont {Kihou}, \citenamefont {Lee},
  \citenamefont {Iyo}, \citenamefont {Eisaki}, \citenamefont {Chuang},
  \citenamefont {Davis},\ and\ \citenamefont {Kim}}]{Nat.Phys.11.177}%
  \BibitemOpen
  \bibfield  {author} {\bibinfo {author} {\bibfnamefont {M.~P.}\ \bibnamefont
  {Allan}}, \bibinfo {author} {\bibfnamefont {K.}~\bibnamefont {Lee}}, \bibinfo
  {author} {\bibfnamefont {A.~W.}\ \bibnamefont {Rost}}, \bibinfo {author}
  {\bibfnamefont {M.~H.}\ \bibnamefont {Fischer}}, \bibinfo {author}
  {\bibfnamefont {F.}~\bibnamefont {Massee}}, \bibinfo {author} {\bibfnamefont
  {K.}~\bibnamefont {Kihou}}, \bibinfo {author} {\bibfnamefont {C.-H.}\
  \bibnamefont {Lee}}, \bibinfo {author} {\bibfnamefont {A.}~\bibnamefont
  {Iyo}}, \bibinfo {author} {\bibfnamefont {H.}~\bibnamefont {Eisaki}},
  \bibinfo {author} {\bibfnamefont {T.-M.}\ \bibnamefont {Chuang}}, \bibinfo
  {author} {\bibfnamefont {J.~C.}\ \bibnamefont {Davis}}, \ and\ \bibinfo
  {author} {\bibfnamefont {E.-A.}\ \bibnamefont {Kim}},\ }\href {\doibase
  10.1038/nphys3187} {\bibfield  {journal} {\bibinfo  {journal} {Nat. Phys.}\
  }\textbf {\bibinfo {volume} {11}},\ \bibinfo {pages} {177} (\bibinfo {year}
  {2015})}\BibitemShut {NoStop}%
\bibitem [{\citenamefont {Umezawa}\ \emph {et~al.}(2012)\citenamefont
  {Umezawa}, \citenamefont {Li}, \citenamefont {Miao}, \citenamefont
  {Nakayama}, \citenamefont {Liu}, \citenamefont {Richard}, \citenamefont
  {Sato}, \citenamefont {He}, \citenamefont {Wang}, \citenamefont {Chen},
  \citenamefont {Ding}, \citenamefont {Takahashi},\ and\ \citenamefont
  {Wang}}]{PhysRevLett.108.037002}%
  \BibitemOpen
  \bibfield  {author} {\bibinfo {author} {\bibfnamefont {K.}~\bibnamefont
  {Umezawa}}, \bibinfo {author} {\bibfnamefont {Y.}~\bibnamefont {Li}},
  \bibinfo {author} {\bibfnamefont {H.}~\bibnamefont {Miao}}, \bibinfo {author}
  {\bibfnamefont {K.}~\bibnamefont {Nakayama}}, \bibinfo {author}
  {\bibfnamefont {Z.-H.}\ \bibnamefont {Liu}}, \bibinfo {author} {\bibfnamefont
  {P.}~\bibnamefont {Richard}}, \bibinfo {author} {\bibfnamefont
  {T.}~\bibnamefont {Sato}}, \bibinfo {author} {\bibfnamefont {J.~B.}\
  \bibnamefont {He}}, \bibinfo {author} {\bibfnamefont {D.-M.}\ \bibnamefont
  {Wang}}, \bibinfo {author} {\bibfnamefont {G.~F.}\ \bibnamefont {Chen}},
  \bibinfo {author} {\bibfnamefont {H.}~\bibnamefont {Ding}}, \bibinfo {author}
  {\bibfnamefont {T.}~\bibnamefont {Takahashi}}, \ and\ \bibinfo {author}
  {\bibfnamefont {S.-C.}\ \bibnamefont {Wang}},\ }\href {\doibase
  10.1103/PhysRevLett.108.037002} {\bibfield  {journal} {\bibinfo  {journal}
  {Phys. Rev. Lett.}\ }\textbf {\bibinfo {volume} {108}},\ \bibinfo {pages}
  {037002} (\bibinfo {year} {2012})}\BibitemShut {NoStop}%
\bibitem [{\citenamefont {Borisenko}\ \emph {et~al.}(2012)\citenamefont
  {Borisenko}, \citenamefont {Zabolotnyy}, \citenamefont {Kordyuk},
  \citenamefont {Evtushinsky}, \citenamefont {Kim}, \citenamefont {Morozov},
  \citenamefont {Follath},\ and\ \citenamefont {Büchner}}]{sym4010251}%
  \BibitemOpen
  \bibfield  {author} {\bibinfo {author} {\bibfnamefont {S.~V.}\ \bibnamefont
  {Borisenko}}, \bibinfo {author} {\bibfnamefont {V.~B.}\ \bibnamefont
  {Zabolotnyy}}, \bibinfo {author} {\bibfnamefont {A.~A.}\ \bibnamefont
  {Kordyuk}}, \bibinfo {author} {\bibfnamefont {D.~V.}\ \bibnamefont
  {Evtushinsky}}, \bibinfo {author} {\bibfnamefont {T.~K.}\ \bibnamefont
  {Kim}}, \bibinfo {author} {\bibfnamefont {I.~V.}\ \bibnamefont {Morozov}},
  \bibinfo {author} {\bibfnamefont {R.}~\bibnamefont {Follath}}, \ and\
  \bibinfo {author} {\bibfnamefont {B.}~\bibnamefont {Büchner}},\ }\href
  {\doibase 10.3390/sym4010251} {\bibfield  {journal} {\bibinfo  {journal}
  {Symmetry}\ }\textbf {\bibinfo {volume} {4}},\ \bibinfo {pages} {251}
  (\bibinfo {year} {2012})}\BibitemShut {NoStop}%
\bibitem [{\citenamefont {Wright}\ \emph {et~al.}(2013)\citenamefont {Wright},
  \citenamefont {Pitcher}, \citenamefont {Trevelyan-Thomas}, \citenamefont
  {Lancaster}, \citenamefont {Baker}, \citenamefont {Pratt}, \citenamefont
  {Clarke},\ and\ \citenamefont {Blundell}}]{PhysRevB.88.060401}%
  \BibitemOpen
  \bibfield  {author} {\bibinfo {author} {\bibfnamefont {J.~D.}\ \bibnamefont
  {Wright}}, \bibinfo {author} {\bibfnamefont {M.~J.}\ \bibnamefont {Pitcher}},
  \bibinfo {author} {\bibfnamefont {W.}~\bibnamefont {Trevelyan-Thomas}},
  \bibinfo {author} {\bibfnamefont {T.}~\bibnamefont {Lancaster}}, \bibinfo
  {author} {\bibfnamefont {P.~J.}\ \bibnamefont {Baker}}, \bibinfo {author}
  {\bibfnamefont {F.~L.}\ \bibnamefont {Pratt}}, \bibinfo {author}
  {\bibfnamefont {S.~J.}\ \bibnamefont {Clarke}}, \ and\ \bibinfo {author}
  {\bibfnamefont {S.~J.}\ \bibnamefont {Blundell}},\ }\href {\doibase
  10.1103/PhysRevB.88.060401} {\bibfield  {journal} {\bibinfo  {journal} {Phys.
  Rev. B}\ }\textbf {\bibinfo {volume} {88}},\ \bibinfo {pages} {060401}
  (\bibinfo {year} {2013})}\BibitemShut {NoStop}%
\bibitem [{\citenamefont {Brand}\ \emph {et~al.}(2014)\citenamefont {Brand},
  \citenamefont {Stunault}, \citenamefont {Wurmehl}, \citenamefont {Harnagea},
  \citenamefont {B\"uchner}, \citenamefont {Meven},\ and\ \citenamefont
  {Braden}}]{PhysRevB.89.045141}%
  \BibitemOpen
  \bibfield  {author} {\bibinfo {author} {\bibfnamefont {J.}~\bibnamefont
  {Brand}}, \bibinfo {author} {\bibfnamefont {A.}~\bibnamefont {Stunault}},
  \bibinfo {author} {\bibfnamefont {S.}~\bibnamefont {Wurmehl}}, \bibinfo
  {author} {\bibfnamefont {L.}~\bibnamefont {Harnagea}}, \bibinfo {author}
  {\bibfnamefont {B.}~\bibnamefont {B\"uchner}}, \bibinfo {author}
  {\bibfnamefont {M.}~\bibnamefont {Meven}}, \ and\ \bibinfo {author}
  {\bibfnamefont {M.}~\bibnamefont {Braden}},\ }\href {\doibase
  10.1103/PhysRevB.89.045141} {\bibfield  {journal} {\bibinfo  {journal} {Phys.
  Rev. B}\ }\textbf {\bibinfo {volume} {89}},\ \bibinfo {pages} {045141}
  (\bibinfo {year} {2014})}\BibitemShut {NoStop}%
\bibitem [{\citenamefont {Wang}\ \emph {et~al.}(2013)\citenamefont {Wang},
  \citenamefont {Kreisel}, \citenamefont {Zabolotnyy}, \citenamefont
  {Borisenko}, \citenamefont {B\"uchner}, \citenamefont {Maier}, \citenamefont
  {Hirschfeld},\ and\ \citenamefont {Scalapino}}]{PhysRevB.88.174516}%
  \BibitemOpen
  \bibfield  {author} {\bibinfo {author} {\bibfnamefont {Y.}~\bibnamefont
  {Wang}}, \bibinfo {author} {\bibfnamefont {A.}~\bibnamefont {Kreisel}},
  \bibinfo {author} {\bibfnamefont {V.~B.}\ \bibnamefont {Zabolotnyy}},
  \bibinfo {author} {\bibfnamefont {S.~V.}\ \bibnamefont {Borisenko}}, \bibinfo
  {author} {\bibfnamefont {B.}~\bibnamefont {B\"uchner}}, \bibinfo {author}
  {\bibfnamefont {T.~A.}\ \bibnamefont {Maier}}, \bibinfo {author}
  {\bibfnamefont {P.~J.}\ \bibnamefont {Hirschfeld}}, \ and\ \bibinfo {author}
  {\bibfnamefont {D.~J.}\ \bibnamefont {Scalapino}},\ }\href {\doibase
  10.1103/PhysRevB.88.174516} {\bibfield  {journal} {\bibinfo  {journal} {Phys.
  Rev. B}\ }\textbf {\bibinfo {volume} {88}},\ \bibinfo {pages} {174516}
  (\bibinfo {year} {2013})}\BibitemShut {NoStop}%
\bibitem [{\citenamefont {Platt}\ \emph {et~al.}(2011)\citenamefont {Platt},
  \citenamefont {Thomale},\ and\ \citenamefont {Hanke}}]{PhysRevB.84.235121}%
  \BibitemOpen
  \bibfield  {author} {\bibinfo {author} {\bibfnamefont {C.}~\bibnamefont
  {Platt}}, \bibinfo {author} {\bibfnamefont {R.}~\bibnamefont {Thomale}}, \
  and\ \bibinfo {author} {\bibfnamefont {W.}~\bibnamefont {Hanke}},\ }\href
  {\doibase 10.1103/PhysRevB.84.235121} {\bibfield  {journal} {\bibinfo
  {journal} {Phys. Rev. B}\ }\textbf {\bibinfo {volume} {84}},\ \bibinfo
  {pages} {235121} (\bibinfo {year} {2011})}\BibitemShut {NoStop}%
\bibitem [{\citenamefont {Ahn}\ \emph {et~al.}(2014)\citenamefont {Ahn},
  \citenamefont {Eremin}, \citenamefont {Knolle}, \citenamefont {Zabolotnyy},
  \citenamefont {Borisenko}, \citenamefont {B\"uchner},\ and\ \citenamefont
  {Chubukov}}]{PhysRevB.89.144513}%
  \BibitemOpen
  \bibfield  {author} {\bibinfo {author} {\bibfnamefont {F.}~\bibnamefont
  {Ahn}}, \bibinfo {author} {\bibfnamefont {I.}~\bibnamefont {Eremin}},
  \bibinfo {author} {\bibfnamefont {J.}~\bibnamefont {Knolle}}, \bibinfo
  {author} {\bibfnamefont {V.~B.}\ \bibnamefont {Zabolotnyy}}, \bibinfo
  {author} {\bibfnamefont {S.~V.}\ \bibnamefont {Borisenko}}, \bibinfo {author}
  {\bibfnamefont {B.}~\bibnamefont {B\"uchner}}, \ and\ \bibinfo {author}
  {\bibfnamefont {A.~V.}\ \bibnamefont {Chubukov}},\ }\href {\doibase
  10.1103/PhysRevB.89.144513} {\bibfield  {journal} {\bibinfo  {journal} {Phys.
  Rev. B}\ }\textbf {\bibinfo {volume} {89}},\ \bibinfo {pages} {144513}
  (\bibinfo {year} {2014})}\BibitemShut {NoStop}%
\bibitem [{\citenamefont {Yin}\ \emph {et~al.}(2014)\citenamefont {Yin},
  \citenamefont {Haule},\ and\ \citenamefont {Kotliar}}]{Nat.Phys.10.845}%
  \BibitemOpen
  \bibfield  {author} {\bibinfo {author} {\bibfnamefont {Z.~P.}\ \bibnamefont
  {Yin}}, \bibinfo {author} {\bibfnamefont {K.}~\bibnamefont {Haule}}, \ and\
  \bibinfo {author} {\bibfnamefont {G.}~\bibnamefont {Kotliar}},\ }\href
  {\doibase 10.1038/nphys3116} {\bibfield  {journal} {\bibinfo  {journal} {Nat.
  Phys.}\ }\textbf {\bibinfo {volume} {10}},\ \bibinfo {pages} {845} (\bibinfo
  {year} {2014})}\BibitemShut {NoStop}%
\bibitem [{\citenamefont {Saito}\ \emph {et~al.}(2014)\citenamefont {Saito},
  \citenamefont {Onari}, \citenamefont {Yamakawa}, \citenamefont {Kontani},
  \citenamefont {Borisenko},\ and\ \citenamefont
  {Zabolotnyy}}]{PhysRevB.90.035104}%
  \BibitemOpen
  \bibfield  {author} {\bibinfo {author} {\bibfnamefont {T.}~\bibnamefont
  {Saito}}, \bibinfo {author} {\bibfnamefont {S.}~\bibnamefont {Onari}},
  \bibinfo {author} {\bibfnamefont {Y.}~\bibnamefont {Yamakawa}}, \bibinfo
  {author} {\bibfnamefont {H.}~\bibnamefont {Kontani}}, \bibinfo {author}
  {\bibfnamefont {S.~V.}\ \bibnamefont {Borisenko}}, \ and\ \bibinfo {author}
  {\bibfnamefont {V.~B.}\ \bibnamefont {Zabolotnyy}},\ }\href {\doibase
  10.1103/PhysRevB.90.035104} {\bibfield  {journal} {\bibinfo  {journal} {Phys.
  Rev. B}\ }\textbf {\bibinfo {volume} {90}},\ \bibinfo {pages} {035104}
  (\bibinfo {year} {2014})}\BibitemShut {NoStop}%
\bibitem [{\citenamefont {Yamada}\ \emph {et~al.}(2014)\citenamefont {Yamada},
  \citenamefont {Ishizuka},\ and\ \citenamefont
  {Ōno}}]{doi:10.7566/JPSJ.83.043704}%
  \BibitemOpen
  \bibfield  {author} {\bibinfo {author} {\bibfnamefont {T.}~\bibnamefont
  {Yamada}}, \bibinfo {author} {\bibfnamefont {J.}~\bibnamefont {Ishizuka}}, \
  and\ \bibinfo {author} {\bibfnamefont {Y.}~\bibnamefont {Ōno}},\ }\href
  {\doibase 10.7566/JPSJ.83.043704} {\bibfield  {journal} {\bibinfo  {journal}
  {Journal of the Physical Society of Japan}\ }\textbf {\bibinfo {volume}
  {83}},\ \bibinfo {pages} {043704} (\bibinfo {year} {2014})}
  \BibitemShut {NoStop}%
\bibitem [{\citenamefont {Lee}\ and\ \citenamefont
  {Wen}(2008)}]{PhysRevB.78.144517}%
  \BibitemOpen
  \bibfield  {author} {\bibinfo {author} {\bibfnamefont {P.~A.}\ \bibnamefont
  {Lee}}\ and\ \bibinfo {author} {\bibfnamefont {X.-G.}\ \bibnamefont {Wen}},\
  }\href {\doibase 10.1103/PhysRevB.78.144517} {\bibfield  {journal} {\bibinfo
  {journal} {Phys. Rev. B}\ }\textbf {\bibinfo {volume} {78}},\ \bibinfo
  {pages} {144517} (\bibinfo {year} {2008})}\BibitemShut {NoStop}%
\bibitem [{\citenamefont {Eschrig}\ and\ \citenamefont
  {Koepernik}(2009)}]{PhysRevB.80.104503}%
  \BibitemOpen
  \bibfield  {author} {\bibinfo {author} {\bibfnamefont {H.}~\bibnamefont
  {Eschrig}}\ and\ \bibinfo {author} {\bibfnamefont {K.}~\bibnamefont
  {Koepernik}},\ }\href {\doibase 10.1103/PhysRevB.80.104503} {\bibfield
  {journal} {\bibinfo  {journal} {Phys. Rev. B}\ }\textbf {\bibinfo {volume}
  {80}},\ \bibinfo {pages} {104503} (\bibinfo {year} {2009})}\BibitemShut
  {NoStop}%
\bibitem [{\citenamefont {Andersen}\ and\ \citenamefont
  {Boeri}(2011)}]{ANDP:ANDP201000149}%
  \BibitemOpen
  \bibfield  {author} {\bibinfo {author} {\bibfnamefont {O.}~\bibnamefont
  {Andersen}}\ and\ \bibinfo {author} {\bibfnamefont {L.}~\bibnamefont
  {Boeri}},\ }\href {\doibase 10.1002/andp.201000149} {\bibfield  {journal}
  {\bibinfo  {journal} {Annalen der Physik}\ }\textbf {\bibinfo {volume}
  {523}},\ \bibinfo {pages} {8} (\bibinfo {year} {2011})}\BibitemShut {NoStop}%
\bibitem [{\citenamefont {Casula}\ and\ \citenamefont
  {Sorella}(2013)}]{PhysRevB.88.155125}%
  \BibitemOpen
  \bibfield  {author} {\bibinfo {author} {\bibfnamefont {M.}~\bibnamefont
  {Casula}}\ and\ \bibinfo {author} {\bibfnamefont {S.}~\bibnamefont
  {Sorella}},\ }\href {\doibase 10.1103/PhysRevB.88.155125} {\bibfield
  {journal} {\bibinfo  {journal} {Phys. Rev. B}\ }\textbf {\bibinfo {volume}
  {88}},\ \bibinfo {pages} {155125} (\bibinfo {year} {2013})}\BibitemShut
  {NoStop}%
\bibitem [{\citenamefont {Fischer}(2013)}]{1367-2630-15-7-073006}%
  \BibitemOpen
  \bibfield  {author} {\bibinfo {author} {\bibfnamefont {M.~H.}\ \bibnamefont
  {Fischer}},\ }\href {http://stacks.iop.org/1367-2630/15/i=7/a=073006}
  {\bibfield  {journal} {\bibinfo  {journal} {New Journal of Physics}\ }\textbf
  {\bibinfo {volume} {15}},\ \bibinfo {pages} {073006} (\bibinfo {year}
  {2013})}\BibitemShut {NoStop}%
\bibitem [{\citenamefont {Cvetkovic}\ and\ \citenamefont
  {Vafek}(2013)}]{PhysRevB.88.134510}%
  \BibitemOpen
  \bibfield  {author} {\bibinfo {author} {\bibfnamefont {V.}~\bibnamefont
  {Cvetkovic}}\ and\ \bibinfo {author} {\bibfnamefont {O.}~\bibnamefont
  {Vafek}},\ }\href {\doibase 10.1103/PhysRevB.88.134510} {\bibfield  {journal}
  {\bibinfo  {journal} {Phys. Rev. B}\ }\textbf {\bibinfo {volume} {88}},\
  \bibinfo {pages} {134510} (\bibinfo {year} {2013})}\BibitemShut {NoStop}%
\bibitem [{\citenamefont {Hu}(2013)}]{PhysRevX.3.031004}%
  \BibitemOpen
  \bibfield  {author} {\bibinfo {author} {\bibfnamefont {J.}~\bibnamefont
  {Hu}},\ }\href {\doibase 10.1103/PhysRevX.3.031004} {\bibfield  {journal}
  {\bibinfo  {journal} {Phys. Rev. X}\ }\textbf {\bibinfo {volume} {3}},\
  \bibinfo {pages} {031004} (\bibinfo {year} {2013})}\BibitemShut {NoStop}%
\bibitem [{\citenamefont {Hao}\ and\ \citenamefont
  {Hu}(2014)}]{PhysRevB.89.045144}%
  \BibitemOpen
  \bibfield  {author} {\bibinfo {author} {\bibfnamefont {N.}~\bibnamefont
  {Hao}}\ and\ \bibinfo {author} {\bibfnamefont {J.}~\bibnamefont {Hu}},\
  }\href {\doibase 10.1103/PhysRevB.89.045144} {\bibfield  {journal} {\bibinfo
  {journal} {Phys. Rev. B}\ }\textbf {\bibinfo {volume} {89}},\ \bibinfo
  {pages} {045144} (\bibinfo {year} {2014})}\BibitemShut {NoStop}%
\bibitem [{\citenamefont {Tzen Ong}, \citenamefont
  {Coleman}\ and\ \citenamefont{Schmalian}(2014)}]{1410.3554}%
  \BibitemOpen
  \bibfield  {author} {\bibinfo {author} {\bibfnamefont {T.}~\bibnamefont
  {Tzen Ong}}, \bibinfo {author} {\bibfnamefont {P.}\ \bibnamefont
  {Coleman}},\ and\ \bibinfo {author} {\bibfnamefont {J.}\ \bibnamefont
  {Schmalian}}\ }\href@noop {} {\bibfield  {journal} {\bibinfo  {journal}
  {ArXiv e-prints}\ } (\bibinfo {year} {2014})},\ \Eprint
  {http://arxiv.org/abs/arXiv:1410.3554} {arXiv:1410.3554} \BibitemShut
  {NoStop}%
\bibitem [{\citenamefont {Wang}\ \emph {et~al.}(2015)\citenamefont {Wang},
  \citenamefont {Berlijn}, \citenamefont {Hirschfeld}, \citenamefont
  {Scalapino},\ and\ \citenamefont {Maier}}]{PhysRevLett.114.107002}%
  \BibitemOpen
  \bibfield  {author} {\bibinfo {author} {\bibfnamefont {Y.}~\bibnamefont
  {Wang}}, \bibinfo {author} {\bibfnamefont {T.}~\bibnamefont {Berlijn}},
  \bibinfo {author} {\bibfnamefont {P.~J.}\ \bibnamefont {Hirschfeld}},
  \bibinfo {author} {\bibfnamefont {D.~J.}\ \bibnamefont {Scalapino}}, \ and\
  \bibinfo {author} {\bibfnamefont {T.~A.}\ \bibnamefont {Maier}},\ }\href
  {\doibase 10.1103/PhysRevLett.114.107002} {\bibfield  {journal} {\bibinfo
  {journal} {Phys. Rev. Lett.}\ }\textbf {\bibinfo {volume} {114}},\ \bibinfo
  {pages} {107002} (\bibinfo {year} {2015})}\BibitemShut {NoStop}%
%
\bibitem [{SM()}]{SM}%
  \BibitemOpen
  \href@noop {} {}\bibinfo {note} {See supplemental material at [], which includes 
  Refs.~\cite{1367-2630-11-2-025016, PhysRevB.75.045118, PhysRevB.86.125114, PhysRevLett.56.2521,
  RevModPhys.83.349, PhysRevB.85.205106, Nat.Commun.4.2783, Sci.Rep.2.381,  0034-4885-74-12-124508, 
  doi:10.1143/JPSJ.79.044705, PhysRevB.85.094509, 1601.05813},  
  for details
  of the electronic structure calculation, the spin fluctuation mediated
  pairing interaction and the RPA susceptibilities.}\BibitemShut {Stop}%
\bibitem [{\citenamefont {Graser}\ \emph {et~al.}(2009)\citenamefont {Graser},
  \citenamefont {Maier}, \citenamefont {Hirschfeld},\ and\ \citenamefont
  {Scalapino}}]{1367-2630-11-2-025016}%
  \BibitemOpen
  \bibfield  {author} {\bibinfo {author} {\bibfnamefont {S.}~\bibnamefont
  {Graser}}, \bibinfo {author} {\bibfnamefont {T.~A.}\ \bibnamefont {Maier}},
  \bibinfo {author} {\bibfnamefont {P.~J.}\ \bibnamefont {Hirschfeld}}, \ and\
  \bibinfo {author} {\bibfnamefont {D.~J.}\ \bibnamefont {Scalapino}},\ }\href
  {http://stacks.iop.org/1367-2630/11/i=2/a=025016} {\bibfield  {journal}
  {\bibinfo  {journal} {New Journal of Physics}\ }\textbf {\bibinfo {volume}
  {11}},\ \bibinfo {pages} {025016} (\bibinfo {year} {2009})}\BibitemShut
  {NoStop}%
\bibitem [{\citenamefont {Toschi}\ \emph {et~al.}(2007)\citenamefont {Toschi},
  \citenamefont {Katanin},\ and\ \citenamefont {Held}}]{PhysRevB.75.045118}%
  \BibitemOpen
  \bibfield  {author} {\bibinfo {author} {\bibfnamefont {A.}~\bibnamefont
  {Toschi}}, \bibinfo {author} {\bibfnamefont {A.~A.}\ \bibnamefont {Katanin}},
  \ and\ \bibinfo {author} {\bibfnamefont {K.}~\bibnamefont {Held}},\ }\href
  {\doibase 10.1103/PhysRevB.75.045118} {\bibfield  {journal} {\bibinfo
  {journal} {Phys. Rev. B}\ }\textbf {\bibinfo {volume} {75}},\ \bibinfo
  {pages} {045118} (\bibinfo {year} {2007})}\BibitemShut {NoStop}%
\bibitem [{\citenamefont {Rohringer}\ \emph {et~al.}(2012)\citenamefont
  {Rohringer}, \citenamefont {Valli},\ and\ \citenamefont
  {Toschi}}]{PhysRevB.86.125114}%
  \BibitemOpen
  \bibfield  {author} {\bibinfo {author} {\bibfnamefont {G.}~\bibnamefont
  {Rohringer}}, \bibinfo {author} {\bibfnamefont {A.}~\bibnamefont {Valli}}, \
  and\ \bibinfo {author} {\bibfnamefont {A.}~\bibnamefont {Toschi}},\ }\href
  {\doibase 10.1103/PhysRevB.86.125114} {\bibfield  {journal} {\bibinfo
  {journal} {Phys. Rev. B}\ }\textbf {\bibinfo {volume} {86}},\ \bibinfo
  {pages} {125114} (\bibinfo {year} {2012})}\BibitemShut {NoStop}%
\bibitem [{\citenamefont {Hirsch}\ and\ \citenamefont
  {Fye}(1986)}]{PhysRevLett.56.2521}%
  \BibitemOpen
  \bibfield  {author} {\bibinfo {author} {\bibfnamefont {J.~E.}\ \bibnamefont
  {Hirsch}}\ and\ \bibinfo {author} {\bibfnamefont {R.~M.}\ \bibnamefont
  {Fye}},\ }\href {\doibase 10.1103/PhysRevLett.56.2521} {\bibfield  {journal}
  {\bibinfo  {journal} {Phys. Rev. Lett.}\ }\textbf {\bibinfo {volume} {56}},\
  \bibinfo {pages} {2521} (\bibinfo {year} {1986})}\BibitemShut {NoStop}%
\bibitem [{\citenamefont {Gull}\ \emph {et~al.}(2011)\citenamefont {Gull},
  \citenamefont {Millis}, \citenamefont {Lichtenstein}, \citenamefont
  {Rubtsov}, \citenamefont {Troyer},\ and\ \citenamefont
  {Werner}}]{RevModPhys.83.349}%
  \BibitemOpen
  \bibfield  {author} {\bibinfo {author} {\bibfnamefont {E.}~\bibnamefont
  {Gull}}, \bibinfo {author} {\bibfnamefont {A.~J.}\ \bibnamefont {Millis}},
  \bibinfo {author} {\bibfnamefont {A.~I.}\ \bibnamefont {Lichtenstein}},
  \bibinfo {author} {\bibfnamefont {A.~N.}\ \bibnamefont {Rubtsov}}, \bibinfo
  {author} {\bibfnamefont {M.}~\bibnamefont {Troyer}}, \ and\ \bibinfo {author}
  {\bibfnamefont {P.}~\bibnamefont {Werner}},\ }\href {\doibase
  10.1103/RevModPhys.83.349} {\bibfield  {journal} {\bibinfo  {journal} {Rev.
  Mod. Phys.}\ }\textbf {\bibinfo {volume} {83}},\ \bibinfo {pages} {349}
  (\bibinfo {year} {2011})}\BibitemShut {NoStop}%
\bibitem [{\citenamefont {Hafermann}\ \emph {et~al.}(2012)\citenamefont
  {Hafermann}, \citenamefont {Patton},\ and\ \citenamefont
  {Werner}}]{PhysRevB.85.205106}%
  \BibitemOpen
  \bibfield  {author} {\bibinfo {author} {\bibfnamefont {H.}~\bibnamefont
  {Hafermann}}, \bibinfo {author} {\bibfnamefont {K.~R.}\ \bibnamefont
  {Patton}}, \ and\ \bibinfo {author} {\bibfnamefont {P.}~\bibnamefont
  {Werner}},\ }\href {\doibase 10.1103/PhysRevB.85.205106} {\bibfield
  {journal} {\bibinfo  {journal} {Phys. Rev. B}\ }\textbf {\bibinfo {volume}
  {85}},\ \bibinfo {pages} {205106} (\bibinfo {year} {2012})}\BibitemShut
  {NoStop}%
\bibitem [{\citenamefont {Yu}\ \emph {et~al.}(2013)\citenamefont {Yu},
  \citenamefont {Goswami}, \citenamefont {Si}, \citenamefont {Nikolic},\ and\
  \citenamefont {Zhu}}]{Nat.Commun.4.2783}%
  \BibitemOpen
  \bibfield  {author} {\bibinfo {author} {\bibfnamefont {R.}~\bibnamefont
  {Yu}}, \bibinfo {author} {\bibfnamefont {P.}~\bibnamefont {Goswami}},
  \bibinfo {author} {\bibfnamefont {Q.}~\bibnamefont {Si}}, \bibinfo {author}
  {\bibfnamefont {P.}~\bibnamefont {Nikolic}}, \ and\ \bibinfo {author}
  {\bibfnamefont {J.~X.}\ \bibnamefont {Zhu}},\ }\href {\doibase
  10.1038/ncomms3783} {\bibfield  {journal} {\bibinfo  {journal} {Nat Commun}\
  }\textbf {\bibinfo {volume} {4}},\ \bibinfo {pages} {2783} (\bibinfo {year}
  {2013})}\BibitemShut {NoStop}%
\bibitem [{\citenamefont {Hu}\ and\ \citenamefont
  {Ding}(2012)}]{Sci.Rep.2.381}%
  \BibitemOpen
  \bibfield  {author} {\bibinfo {author} {\bibfnamefont {J.}~\bibnamefont
  {Hu}}\ and\ \bibinfo {author} {\bibfnamefont {H.}~\bibnamefont {Ding}},\
  }\href {\doibase 10.1038/srep00381} {\bibfield  {journal} {\bibinfo
  {journal} {Sci. Rep.}\ }\textbf {\bibinfo {volume} {2}},\ \bibinfo {pages}
  {381} (\bibinfo {year} {2012})}\BibitemShut {NoStop}%
\bibitem [{\citenamefont {Hirschfeld}\ \emph {et~al.}(2011)\citenamefont
  {Hirschfeld}, \citenamefont {Korshunov},\ and\ \citenamefont
  {Mazin}}]{0034-4885-74-12-124508}%
  \BibitemOpen
  \bibfield  {author} {\bibinfo {author} {\bibfnamefont {P.~J.}\ \bibnamefont
  {Hirschfeld}}, \bibinfo {author} {\bibfnamefont {M.~M.}\ \bibnamefont
  {Korshunov}}, \ and\ \bibinfo {author} {\bibfnamefont {I.~I.}\ \bibnamefont
  {Mazin}},\ }\href {http://stacks.iop.org/0034-4885/74/i=12/a=124508}
  {\bibfield  {journal} {\bibinfo  {journal} {Reports on Progress in Physics}\
  }\textbf {\bibinfo {volume} {74}},\ \bibinfo {pages} {124508} (\bibinfo
  {year} {2011})}\BibitemShut {NoStop}%
\bibitem [{\citenamefont {Miyake}\ \emph {et~al.}(2010)\citenamefont {Miyake},
  \citenamefont {Nakamura}, \citenamefont {Arita},\ and\ \citenamefont
  {Imada}}]{doi:10.1143/JPSJ.79.044705}%
  \BibitemOpen
  \bibfield  {author} {\bibinfo {author} {\bibfnamefont {T.}~\bibnamefont
  {Miyake}}, \bibinfo {author} {\bibfnamefont {K.}~\bibnamefont {Nakamura}},
  \bibinfo {author} {\bibfnamefont {R.}~\bibnamefont {Arita}}, \ and\ \bibinfo
  {author} {\bibfnamefont {M.}~\bibnamefont {Imada}},\ }\href {\doibase
  10.1143/JPSJ.79.044705} {\bibfield  {journal} {\bibinfo  {journal} {Journal
  of the Physical Society of Japan}\ }\textbf {\bibinfo {volume} {79}},\
  \bibinfo {pages} {044705} (\bibinfo {year} {2010})} \BibitemShut
  {NoStop}%
\bibitem [{\citenamefont {Hajiri}\ \emph {et~al.}(2012)\citenamefont {Hajiri},
  \citenamefont {Ito}, \citenamefont {Niwa}, \citenamefont {Matsunami},
  \citenamefont {Min}, \citenamefont {Kwon},\ and\ \citenamefont
  {Kimura}}]{PhysRevB.85.094509}%
  \BibitemOpen
  \bibfield  {author} {\bibinfo {author} {\bibfnamefont {T.}~\bibnamefont
  {Hajiri}}, \bibinfo {author} {\bibfnamefont {T.}~\bibnamefont {Ito}},
  \bibinfo {author} {\bibfnamefont {R.}~\bibnamefont {Niwa}}, \bibinfo {author}
  {\bibfnamefont {M.}~\bibnamefont {Matsunami}}, \bibinfo {author}
  {\bibfnamefont {B.~H.}\ \bibnamefont {Min}}, \bibinfo {author} {\bibfnamefont
  {Y.~S.}\ \bibnamefont {Kwon}}, \ and\ \bibinfo {author} {\bibfnamefont
  {S.}~\bibnamefont {Kimura}},\ }\href {\doibase 10.1103/PhysRevB.85.094509}
  {\bibfield  {journal} {\bibinfo  {journal} {Phys. Rev. B}\ }\textbf {\bibinfo
  {volume} {85}},\ \bibinfo {pages} {094509} (\bibinfo {year}
  {2012})}\BibitemShut {NoStop}%
\bibitem [{\citenamefont {Nourafkan}\ and\ \citenamefont
  {Tremblay}(2016)}]{1601.05813}%
  \BibitemOpen
  \bibfield  {author} {\bibinfo {author} {\bibfnamefont {R.}~\bibnamefont
  {Nourafkan}}\ and\ \bibinfo {author} {\bibfnamefont {A.-M.}\ \bibnamefont
  {Tremblay}},\ }\href@noop {} {\bibfield  {journal} {\bibinfo  {journal}
  {ArXiv e-prints}\ } (\bibinfo {year} {2016})},\ \Eprint
  {http://arxiv.org/abs/arXiv:1601.05813} {arXiv:1601.05813} \BibitemShut
  {NoStop}%
\bibitem [{\citenamefont {Miao}\ \emph {et~al.}(2015)\citenamefont {Miao},
  \citenamefont {Qian}, \citenamefont {Shi}, \citenamefont {Richard},
  \citenamefont {Kim}, \citenamefont {Hoesch}, \citenamefont {Xing},
  \citenamefont {Wang}, \citenamefont {Jin}, \citenamefont {Hu},\ and\
  \citenamefont {Ding}}]{Nat.Commun.6.7056}%
  \BibitemOpen
  \bibfield  {author} {\bibinfo {author} {\bibfnamefont {H.}~\bibnamefont
  {Miao}}, \bibinfo {author} {\bibfnamefont {T.}~\bibnamefont {Qian}}, \bibinfo
  {author} {\bibfnamefont {X.}~\bibnamefont {Shi}}, \bibinfo {author}
  {\bibfnamefont {P.}~\bibnamefont {Richard}}, \bibinfo {author} {\bibfnamefont
  {T.~K.}\ \bibnamefont {Kim}}, \bibinfo {author} {\bibfnamefont
  {M.}~\bibnamefont {Hoesch}}, \bibinfo {author} {\bibfnamefont {L.~Y.}\
  \bibnamefont {Xing}}, \bibinfo {author} {\bibfnamefont {X.-C.}\ \bibnamefont
  {Wang}}, \bibinfo {author} {\bibfnamefont {C.-Q.}\ \bibnamefont {Jin}},
  \bibinfo {author} {\bibfnamefont {J.-P.}\ \bibnamefont {Hu}}, \ and\ \bibinfo
  {author} {\bibfnamefont {H.}~\bibnamefont {Ding}},\ }\href {\doibase
  10.1038/ncomms7056} {\bibfield  {journal} {\bibinfo  {journal} {Nat Commun}\
  }\textbf {\bibinfo {volume} {6}},\ \bibinfo {pages} {7056} (\bibinfo {year}
  {2015})}\BibitemShut {NoStop}%
\bibitem [{\citenamefont {Aichhorn}\ \emph {et~al.}(2009)\citenamefont
  {Aichhorn}, \citenamefont {Pourovskii}, \citenamefont {Vildosola},
  \citenamefont {Ferrero}, \citenamefont {Parcollet}, \citenamefont {Miyake},
  \citenamefont {Georges},\ and\ \citenamefont
  {Biermann}}]{PhysRevB.80.085101}%
  \BibitemOpen
  \bibfield  {author} {\bibinfo {author} {\bibfnamefont {M.}~\bibnamefont
  {Aichhorn}}, \bibinfo {author} {\bibfnamefont {L.}~\bibnamefont
  {Pourovskii}}, \bibinfo {author} {\bibfnamefont {V.}~\bibnamefont
  {Vildosola}}, \bibinfo {author} {\bibfnamefont {M.}~\bibnamefont {Ferrero}},
  \bibinfo {author} {\bibfnamefont {O.}~\bibnamefont {Parcollet}}, \bibinfo
  {author} {\bibfnamefont {T.}~\bibnamefont {Miyake}}, \bibinfo {author}
  {\bibfnamefont {A.}~\bibnamefont {Georges}}, \ and\ \bibinfo {author}
  {\bibfnamefont {S.}~\bibnamefont {Biermann}},\ }\href {\doibase
  10.1103/PhysRevB.80.085101} {\bibfield  {journal} {\bibinfo  {journal} {Phys.
  Rev. B}\ }\textbf {\bibinfo {volume} {80}},\ \bibinfo {pages} {085101}
  (\bibinfo {year} {2009})}\BibitemShut {NoStop}%
\bibitem [{\citenamefont {Haule}\ \emph {et~al.}(2010)\citenamefont {Haule},
  \citenamefont {Yee},\ and\ \citenamefont {Kim}}]{PhysRevB.81.195107}%
  \BibitemOpen
  \bibfield  {author} {\bibinfo {author} {\bibfnamefont {K.}~\bibnamefont
  {Haule}}, \bibinfo {author} {\bibfnamefont {C.-H.}\ \bibnamefont {Yee}}, \
  and\ \bibinfo {author} {\bibfnamefont {K.}~\bibnamefont {Kim}},\ }\href
  {\doibase 10.1103/PhysRevB.81.195107} {\bibfield  {journal} {\bibinfo
  {journal} {Phys. Rev. B}\ }\textbf {\bibinfo {volume} {81}},\ \bibinfo
  {pages} {195107} (\bibinfo {year} {2010})}\BibitemShut {NoStop}%
\bibitem [{\citenamefont {Yin}\ \emph {et~al.}(2011)\citenamefont {Yin},
  \citenamefont {Haule},\ and\ \citenamefont {Kotliar}}]{Nat.Mater.10.1038}%
  \BibitemOpen
  \bibfield  {author} {\bibinfo {author} {\bibfnamefont {Z.~P.}\ \bibnamefont
  {Yin}}, \bibinfo {author} {\bibfnamefont {K.}~\bibnamefont {Haule}}, \ and\
  \bibinfo {author} {\bibfnamefont {G.}~\bibnamefont {Kotliar}},\ }\href
  {\doibase 10.1038/nmat3120} {\bibfield  {journal} {\bibinfo  {journal} {Nat.
  Mater.}\ }\textbf {\bibinfo {volume} {10}},\ \bibinfo {pages} {932} (\bibinfo
  {year} {2011})}\BibitemShut {NoStop}%
\bibitem [{Note1()}]{Note1}%
  \BibitemOpen
  \bibinfo {note} {To keep a minimum variation of orbital content within a
  pocket, the electron pockets are chosen as inner and outer pockets $\beta _1$
  and $\beta _2$ rather than as two crossed ellipse-like pockets of equal
  size.}\BibitemShut {Stop}%
\bibitem [{\citenamefont {Ferber}\ \emph {et~al.}(2012)\citenamefont {Ferber},
  \citenamefont {Foyevtsova}, \citenamefont {Valent\'\i},\ and\ \citenamefont
  {Jeschke}}]{PhysRevB.85.094505}%
  \BibitemOpen
  \bibfield  {author} {\bibinfo {author} {\bibfnamefont {J.}~\bibnamefont
  {Ferber}}, \bibinfo {author} {\bibfnamefont {K.}~\bibnamefont {Foyevtsova}},
  \bibinfo {author} {\bibfnamefont {R.}~\bibnamefont {Valent\'\i}}, \ and\
  \bibinfo {author} {\bibfnamefont {H.~O.}\ \bibnamefont {Jeschke}},\ }\href
  {\doibase 10.1103/PhysRevB.85.094505} {\bibfield  {journal} {\bibinfo
  {journal} {Phys. Rev. B}\ }\textbf {\bibinfo {volume} {85}},\ \bibinfo
  {pages} {094505} (\bibinfo {year} {2012})}\BibitemShut {NoStop}%
\bibitem [{\citenamefont {Bickers}(2004)}]{Bickers2004}%
  \BibitemOpen
  \bibfield  {author} {\bibinfo {author} {\bibfnamefont {N.~E.}\ \bibnamefont
  {Bickers}},\ }\enquote {\bibinfo {title} {Self-consistent many-body theory
  for condensed matter systems},}\ \ (\bibinfo  {publisher} {Springer-Verlag},\
  \bibinfo {address} {New York},\ \bibinfo {year} {2004})\ Chap.~\bibinfo
  {chapter} {6}, pp.\ \bibinfo {pages} {237--296}\BibitemShut {NoStop}%
\bibitem [{\citenamefont {Esirgen}\ and\ \citenamefont
  {Bickers}(1998)}]{PhysRevB.57.5376}%
  \BibitemOpen
  \bibfield  {author} {\bibinfo {author} {\bibfnamefont {G.}~\bibnamefont
  {Esirgen}}\ and\ \bibinfo {author} {\bibfnamefont {N.~E.}\ \bibnamefont
  {Bickers}},\ }\href {\doibase 10.1103/PhysRevB.57.5376} {\bibfield  {journal}
  {\bibinfo  {journal} {Phys. Rev. B}\ }\textbf {\bibinfo {volume} {57}},\
  \bibinfo {pages} {5376} (\bibinfo {year} {1998})}\BibitemShut {NoStop}%
\bibitem [{\citenamefont {Toschi}\ \emph {et~al.}(2012)\citenamefont {Toschi},
  \citenamefont {Arita}, \citenamefont {Hansmann}, \citenamefont
  {Sangiovanni},\ and\ \citenamefont {Held}}]{PhysRevB.86.064411}%
  \BibitemOpen
  \bibfield  {author} {\bibinfo {author} {\bibfnamefont {A.}~\bibnamefont
  {Toschi}}, \bibinfo {author} {\bibfnamefont {R.}~\bibnamefont {Arita}},
  \bibinfo {author} {\bibfnamefont {P.}~\bibnamefont {Hansmann}}, \bibinfo
  {author} {\bibfnamefont {G.}~\bibnamefont {Sangiovanni}}, \ and\ \bibinfo
  {author} {\bibfnamefont {K.}~\bibnamefont {Held}},\ }\href {\doibase
  10.1103/PhysRevB.86.064411} {\bibfield  {journal} {\bibinfo  {journal} {Phys.
  Rev. B}\ }\textbf {\bibinfo {volume} {86}},\ \bibinfo {pages} {064411}
  (\bibinfo {year} {2012})}\BibitemShut {NoStop}%
\bibitem [{\citenamefont {Maier}\ \emph {et~al.}(2007)\citenamefont {Maier},
  \citenamefont {Jarrell},\ and\ \citenamefont
  {Scalapino}}]{PhysRevB.75.134519}%
  \BibitemOpen
  \bibfield  {author} {\bibinfo {author} {\bibfnamefont {T.~A.}\ \bibnamefont
  {Maier}}, \bibinfo {author} {\bibfnamefont {M.}~\bibnamefont {Jarrell}}, \
  and\ \bibinfo {author} {\bibfnamefont {D.~J.}\ \bibnamefont {Scalapino}},\
  }\href {\doibase 10.1103/PhysRevB.75.134519} {\bibfield  {journal} {\bibinfo
  {journal} {Phys. Rev. B}\ }\textbf {\bibinfo {volume} {75}},\ \bibinfo
  {pages} {134519} (\bibinfo {year} {2007})}\BibitemShut {NoStop}%
\bibitem [{\citenamefont {Yanagi}\ \emph {et~al.}(2010)\citenamefont {Yanagi},
  \citenamefont {Yamakawa},\ and\ \citenamefont {\ifmmode~\bar{O}\else
  \={O}\fi{}no}}]{PhysRevB.81.054518}%
  \BibitemOpen
  \bibfield  {author} {\bibinfo {author} {\bibfnamefont {Y.}~\bibnamefont
  {Yanagi}}, \bibinfo {author} {\bibfnamefont {Y.}~\bibnamefont {Yamakawa}}, \
  and\ \bibinfo {author} {\bibfnamefont {Y.}~\bibnamefont
  {\ifmmode~\bar{O}\else \={O}\fi{}no}},\ }\href {\doibase
  10.1103/PhysRevB.81.054518} {\bibfield  {journal} {\bibinfo  {journal} {Phys.
  Rev. B}\ }\textbf {\bibinfo {volume} {81}},\ \bibinfo {pages} {054518}
  (\bibinfo {year} {2010})}\BibitemShut {NoStop}%
\bibitem [{\citenamefont {Miyahara}\ \emph {et~al.}(2013)\citenamefont
  {Miyahara}, \citenamefont {Arita},\ and\ \citenamefont
  {Ikeda}}]{PhysRevB.87.045113}%
  \BibitemOpen
  \bibfield  {author} {\bibinfo {author} {\bibfnamefont {H.}~\bibnamefont
  {Miyahara}}, \bibinfo {author} {\bibfnamefont {R.}~\bibnamefont {Arita}}, \
  and\ \bibinfo {author} {\bibfnamefont {H.}~\bibnamefont {Ikeda}},\ }\href
  {\doibase 10.1103/PhysRevB.87.045113} {\bibfield  {journal} {\bibinfo
  {journal} {Phys. Rev. B}\ }\textbf {\bibinfo {volume} {87}},\ \bibinfo
  {pages} {045113} (\bibinfo {year} {2013})}\BibitemShut {NoStop}%
\bibitem [{Note2()}]{Note2}%
  \BibitemOpen
  \bibinfo {note} {The distance from magnetic and charge/orbital fluctuation
  criticality is determined by the corresponding (dimensionless) magnetic
  (density) Stoner factor $\alpha ^{m(d)}_{\protect \bf q}$, which is the
  largest eigenvalue of ${\protect \bm {\Gamma }}^{irr, m}{\protect \bm {\chi
  }}_{ph}^0({\protect \bf q}, i\nu _n=0)$ ($-{\protect \bm {\Gamma }}^{irr,
  d}{\protect \bm {\chi }}_{ph}^0({\protect \bf q}, i\nu _n=0)$).}\BibitemShut
  {Stop}%
\bibitem [{Note3()}]{Note3}%
  \BibitemOpen
  \bibinfo {note} {The $22;44$ ($44;55$) components in the magnetic and charge
  susceptibilities in the p-h channel are related to the $24;42$ and $42;42$
  ($45;54$ and $ 54;54$) components of the pairing interaction in the p-p
  channel.}\BibitemShut {Stop}%
\bibitem [{\citenamefont {Nourafkan}(2016)}]{PhysRevB.93.241116}%
  \BibitemOpen
  \bibfield  {author} {\bibinfo {author} {\bibfnamefont {R.}~\bibnamefont
  {Nourafkan}},\ }\href {\doibase 10.1103/PhysRevB.93.241116} {\bibfield
  {journal} {\bibinfo  {journal} {Phys. Rev. B}\ }\textbf {\bibinfo {volume}
  {93}},\ \bibinfo {pages} {241116} (\bibinfo {year} {2016})}\BibitemShut
  {NoStop}%
\bibitem [{Note6()}]{Note6}%
  \BibitemOpen
  \bibinfo {note} {The combination of these relations gives $\Delta^{AA(BB)}_{l_1l_2}({\bf k},i\omega_m)=\Delta^{BB(AA)}_{l_2l_1}({\bf k},-i\omega_m)$.}\BibitemShut {Stop}%
\bibitem [{Note7()}]{Note7}%
  \BibitemOpen
  \bibinfo {note} {The five Fe $3d$ orbitals can be categorized into even
  orbital parity ($d_{3z^2}$, $d_{x^2-y^2}$ , $d_{xy}$) with $p_l = +1$ and odd
  orbital parity ($d_{xz}$, $d_{yz}$) with $p_l = -1$.}\BibitemShut {Stop}%
\bibitem [{Note8()}]{Note8}%
  \BibitemOpen
  \bibinfo {note} {The linearized Eliashberg gap equation only gives gap
  symmetry, not gap magnitude. But to make contact with experiment one can
  approximately extract the relative size of the gaps on the different FSs.
  This can be done by defining a Bogoliubov quasi-particle Hamiltonian
  including the real part of the self-energy at the Fermi level in the normal
  part, and employing the gap function obtained from the gap equation as an
  estimate of the anomalous self-energy.~\cite {Nat.Phys.10.845} After
  diagonalizing the Bogoliubov quasi-particle Hamiltonian, the gap magnitude at
  momentum $\protect \bf k$ is given by half of the difference between the
  smallest positive eigenvalue and the largest negative eigenvalue. This is the
  quasi-particle gap which reduces to the SC gap on the FSs. For this
  calculation, the gap function on a very dense $\protect \bf k$-mesh is
  required. Since the gap function is a smooth function, its magnitude on a
  denser mesh can be obtained by spline interpolation.}\BibitemShut {Stop}%
\bibitem [{\citenamefont {Borisenko}\ \emph {et~al.}(2014)\citenamefont
  {Borisenko}, \citenamefont {Evtushinsky}, \citenamefont {Morozov},
  \citenamefont {Wurmehl}, \citenamefont {Büchner}, \citenamefont {Yaresko},
  \citenamefont {Kim}, \citenamefont {Hoesch}, \citenamefont {Wolf},\ and\
  \citenamefont {Zhigadlo}}]{1409.8669}%
  \BibitemOpen
  \bibfield  {author} {\bibinfo {author} {\bibfnamefont {S.}~\bibnamefont
  {Borisenko}}, \bibinfo {author} {\bibfnamefont {D.}~\bibnamefont
  {Evtushinsky}}, \bibinfo {author} {\bibfnamefont {I.}~\bibnamefont
  {Morozov}}, \bibinfo {author} {\bibfnamefont {S.}~\bibnamefont {Wurmehl}},
  \bibinfo {author} {\bibfnamefont {B.}~\bibnamefont {Büchner}}, \bibinfo
  {author} {\bibfnamefont {A.}~\bibnamefont {Yaresko}}, \bibinfo {author}
  {\bibfnamefont {T.}~\bibnamefont {Kim}}, \bibinfo {author} {\bibfnamefont
  {M.}~\bibnamefont {Hoesch}}, \bibinfo {author} {\bibfnamefont
  {T.}~\bibnamefont {Wolf}}, \ and\ \bibinfo {author} {\bibfnamefont
  {N.}~\bibnamefont {Zhigadlo}},\ }\href@noop {} {\enquote {\bibinfo {title}
  {Direct observation of spin-orbit coupling in iron-based superconductors},}\
  } (\bibinfo {year} {2014}),\ \Eprint {http://arxiv.org/abs/arXiv:1409.8669}
  {arXiv:1409.8669} \BibitemShut {NoStop}%
\bibitem [{\citenamefont {Nag}, \citenamefont
  {Schlegel}, \citenamefont {Baumann}, \citenamefont {Grafe}, \citenamefont
  {Beck}, \citenamefont
  {Wurmehl}, \citenamefont
  {Buchner}\ and\ \citenamefont{Hess}(2015)}]{1509.03431}%
  \BibitemOpen
  \bibfield  {author} {\bibinfo {author} {\bibfnamefont {P. T.}~\bibnamefont
  {Nag}}, \bibinfo {author} {\bibfnamefont {R.}\ \bibnamefont
  {Schlegel}},\bibinfo {author} {\bibfnamefont {D.}\ \bibnamefont
  {Baumann}},\bibinfo {author} {\bibfnamefont {H.-J}\ \bibnamefont
  {Grafe}},\bibinfo {author} {\bibfnamefont {R.}\ \bibnamefont
  {Beck}},\bibinfo {author} {\bibfnamefont {S.}\ \bibnamefont
  {Wurmehl}},\bibinfo {author} {\bibfnamefont {B.}\ \bibnamefont
  {Buchner}},\ and\ \bibinfo {author} {\bibfnamefont {C.}\ \bibnamefont
  {Hess}}\ }\href@noop {} {\bibfield  {journal} {\bibinfo  {journal}
  {ArXiv e-prints}\ } (\bibinfo {year} {2015})},\ \Eprint
  {http://arxiv.org/abs/arXiv:1509.03431} {arXiv:1509.03431} \BibitemShut
  {NoStop}%
\end{thebibliography}

\begin{thebibliography}{24}%
\makeatletter
\providecommand \@ifxundefined [1]{%
 \@ifx{#1\undefined}
}%
\providecommand \@ifnum [1]{%
 \ifnum #1\expandafter \@firstoftwo
 \else \expandafter \@secondoftwo
 \fi
}%
\providecommand \@ifx [1]{%
 \ifx #1\expandafter \@firstoftwo
 \else \expandafter \@secondoftwo
 \fi
}%
\providecommand \natexlab [1]{#1}%
\providecommand \enquote  [1]{``#1''}%
\providecommand \bibnamefont  [1]{#1}%
\providecommand \bibfnamefont [1]{#1}%
\providecommand \citenamefont [1]{#1}%
\providecommand \href@noop [0]{\@secondoftwo}%
\providecommand \href [0]{\begingroup \@sanitize@url \@href}%
\providecommand \@href[1]{\@@startlink{#1}\@@href}%
\providecommand \@@href[1]{\endgroup#1\@@endlink}%
\providecommand \@sanitize@url [0]{\catcode `\\12\catcode `\$12\catcode
  `\&12\catcode `\#12\catcode `\^12\catcode `\_12\catcode `\%12\relax}%
\providecommand \@@startlink[1]{}%
\providecommand \@@endlink[0]{}%
\providecommand \url  [0]{\begingroup\@sanitize@url \@url }%
\providecommand \@url [1]{\endgroup\@href {#1}{\urlprefix }}%
\providecommand \urlprefix  [0]{URL }%
\providecommand \Eprint [0]{\href }%
\providecommand \doibase [0]{http://dx.doi.org/}%
\providecommand \selectlanguage [0]{\@gobble}%
\providecommand \bibinfo  [0]{\@secondoftwo}%
\providecommand \bibfield  [0]{\@secondoftwo}%
\providecommand \translation [1]{[#1]}%
\providecommand \BibitemOpen [0]{}%
\providecommand \bibitemStop [0]{}%
\providecommand \bibitemNoStop [0]{.\EOS\space}%
\providecommand \EOS [0]{\spacefactor3000\relax}%
\providecommand \BibitemShut  [1]{\csname bibitem#1\endcsname}%
\let\auto@bib@innerbib\@empty
\bibitem [{\citenamefont {Haule}\ \emph {et~al.}(2010)\citenamefont {Haule},
  \citenamefont {Yee},\ and\ \citenamefont {Kim}}]{PhysRevB.81.195107-SM}%
  \BibitemOpen
  \bibfield  {author} {\bibinfo {author} {\bibfnamefont {K.}~\bibnamefont
  {Haule}}, \bibinfo {author} {\bibfnamefont {C.-H.}\ \bibnamefont {Yee}}, \
  and\ \bibinfo {author} {\bibfnamefont {K.}~\bibnamefont {Kim}},\ }\href
  {\doibase 10.1103/PhysRevB.81.195107} {\bibfield  {journal} {\bibinfo
  {journal} {Phys. Rev. B}\ }\textbf {\bibinfo {volume} {81}},\ \bibinfo
  {pages} {195107} (\bibinfo {year} {2010})}\BibitemShut {NoStop}%
\bibitem [{\citenamefont {Ferber}\ \emph {et~al.}(2012)\citenamefont {Ferber},
  \citenamefont {Foyevtsova}, \citenamefont {Valent\'\i},\ and\ \citenamefont
  {Jeschke}}]{PhysRevB.85.094505-SM}%
  \BibitemOpen
  \bibfield  {author} {\bibinfo {author} {\bibfnamefont {J.}~\bibnamefont
  {Ferber}}, \bibinfo {author} {\bibfnamefont {K.}~\bibnamefont {Foyevtsova}},
  \bibinfo {author} {\bibfnamefont {R.}~\bibnamefont {Valent\'\i}}, \ and\
  \bibinfo {author} {\bibfnamefont {H.~O.}\ \bibnamefont {Jeschke}},\ }\href
  {\doibase 10.1103/PhysRevB.85.094505} {\bibfield  {journal} {\bibinfo
  {journal} {Phys. Rev. B}\ }\textbf {\bibinfo {volume} {85}},\ \bibinfo
  {pages} {094505} (\bibinfo {year} {2012})}\BibitemShut {NoStop}%
\bibitem [{\citenamefont {Borisenko}\ \emph {et~al.}(2010)\citenamefont
  {Borisenko}, \citenamefont {Zabolotnyy}, \citenamefont {Evtushinsky},
  \citenamefont {Kim}, \citenamefont {Morozov}, \citenamefont {Yaresko},
  \citenamefont {Kordyuk}, \citenamefont {Behr}, \citenamefont {Vasiliev},
  \citenamefont {Follath},\ and\ \citenamefont
  {B\"uchner}}]{PhysRevLett.105.067002-SM}%
  \BibitemOpen
  \bibfield  {author} {\bibinfo {author} {\bibfnamefont {S.~V.}\ \bibnamefont
  {Borisenko}}, \bibinfo {author} {\bibfnamefont {V.~B.}\ \bibnamefont
  {Zabolotnyy}}, \bibinfo {author} {\bibfnamefont {D.~V.}\ \bibnamefont
  {Evtushinsky}}, \bibinfo {author} {\bibfnamefont {T.~K.}\ \bibnamefont
  {Kim}}, \bibinfo {author} {\bibfnamefont {I.~V.}\ \bibnamefont {Morozov}},
  \bibinfo {author} {\bibfnamefont {A.~N.}\ \bibnamefont {Yaresko}}, \bibinfo
  {author} {\bibfnamefont {A.~A.}\ \bibnamefont {Kordyuk}}, \bibinfo {author}
  {\bibfnamefont {G.}~\bibnamefont {Behr}}, \bibinfo {author} {\bibfnamefont
  {A.}~\bibnamefont {Vasiliev}}, \bibinfo {author} {\bibfnamefont
  {R.}~\bibnamefont {Follath}}, \ and\ \bibinfo {author} {\bibfnamefont
  {B.}~\bibnamefont {B\"uchner}},\ }\href {\doibase
  10.1103/PhysRevLett.105.067002} {\bibfield  {journal} {\bibinfo  {journal}
  {Phys. Rev. Lett.}\ }\textbf {\bibinfo {volume} {105}},\ \bibinfo {pages}
  {067002} (\bibinfo {year} {2010})}\BibitemShut {NoStop}%
\bibitem [{\citenamefont {Brydon}\ \emph {et~al.}(2011)\citenamefont {Brydon},
  \citenamefont {Daghofer}, \citenamefont {Timm},\ and\ \citenamefont {van~den
  Brink}}]{PhysRevB.83.060501-SM}%
  \BibitemOpen
  \bibfield  {author} {\bibinfo {author} {\bibfnamefont {P.~M.~R.}\
  \bibnamefont {Brydon}}, \bibinfo {author} {\bibfnamefont {M.}~\bibnamefont
  {Daghofer}}, \bibinfo {author} {\bibfnamefont {C.}~\bibnamefont {Timm}}, \
  and\ \bibinfo {author} {\bibfnamefont {J.}~\bibnamefont {van~den Brink}},\
  }\href {\doibase 10.1103/PhysRevB.83.060501} {\bibfield  {journal} {\bibinfo
  {journal} {Phys. Rev. B}\ }\textbf {\bibinfo {volume} {83}},\ \bibinfo
  {pages} {060501} (\bibinfo {year} {2011})}\BibitemShut {NoStop}%
\bibitem [{\citenamefont {Cvetkovic}\ and\ \citenamefont
  {Vafek}(2013)}]{PhysRevB.88.134510-SM}%
  \BibitemOpen
  \bibfield  {author} {\bibinfo {author} {\bibfnamefont {V.}~\bibnamefont
  {Cvetkovic}}\ and\ \bibinfo {author} {\bibfnamefont {O.}~\bibnamefont
  {Vafek}},\ }\href {\doibase 10.1103/PhysRevB.88.134510} {\bibfield  {journal}
  {\bibinfo  {journal} {Phys. Rev. B}\ }\textbf {\bibinfo {volume} {88}},\
  \bibinfo {pages} {134510} (\bibinfo {year} {2013})}\BibitemShut {NoStop}%
\bibitem [{\citenamefont {Bickers}(2004)}]{Bickers2004-SM}%
  \BibitemOpen
  \bibfield  {author} {\bibinfo {author} {\bibfnamefont {N.~E.}\ \bibnamefont
  {Bickers}},\ }\enquote {\bibinfo {title} {Self-consistent many-body theory
  for condensed matter systems},}\ \ (\bibinfo  {publisher} {Springer-Verlag},\
  \bibinfo {address} {New York},\ \bibinfo {year} {2004})\ Chap.~\bibinfo
  {chapter} {6}, pp.\ \bibinfo {pages} {237--296}\BibitemShut {NoStop}%
\bibitem [{\citenamefont {Esirgen}\ and\ \citenamefont
  {Bickers}(1998)}]{PhysRevB.57.5376-SM}%
  \BibitemOpen
  \bibfield  {author} {\bibinfo {author} {\bibfnamefont {G.}~\bibnamefont
  {Esirgen}}\ and\ \bibinfo {author} {\bibfnamefont {N.~E.}\ \bibnamefont
  {Bickers}},\ }\href {\doibase 10.1103/PhysRevB.57.5376} {\bibfield  {journal}
  {\bibinfo  {journal} {Phys. Rev. B}\ }\textbf {\bibinfo {volume} {57}},\
  \bibinfo {pages} {5376} (\bibinfo {year} {1998})}\BibitemShut {NoStop}%
\bibitem [{\citenamefont {Graser}\ \emph {et~al.}(2009)\citenamefont {Graser},
  \citenamefont {Maier}, \citenamefont {Hirschfeld},\ and\ \citenamefont
  {Scalapino}}]{1367-2630-11-2-025016-SM}%
  \BibitemOpen
  \bibfield  {author} {\bibinfo {author} {\bibfnamefont {S.}~\bibnamefont
  {Graser}}, \bibinfo {author} {\bibfnamefont {T.~A.}\ \bibnamefont {Maier}},
  \bibinfo {author} {\bibfnamefont {P.~J.}\ \bibnamefont {Hirschfeld}}, \ and\
  \bibinfo {author} {\bibfnamefont {D.~J.}\ \bibnamefont {Scalapino}},\ }\href
  {http://stacks.iop.org/1367-2630/11/i=2/a=025016} {\bibfield  {journal}
  {\bibinfo  {journal} {New Journal of Physics}\ }\textbf {\bibinfo {volume}
  {11}},\ \bibinfo {pages} {025016} (\bibinfo {year} {2009})}\BibitemShut
  {NoStop}%
\bibitem [{\citenamefont {Toschi}\ \emph {et~al.}(2007)\citenamefont {Toschi},
  \citenamefont {Katanin},\ and\ \citenamefont {Held}}]{PhysRevB.75.045118-SM}%
  \BibitemOpen
  \bibfield  {author} {\bibinfo {author} {\bibfnamefont {A.}~\bibnamefont
  {Toschi}}, \bibinfo {author} {\bibfnamefont {A.~A.}\ \bibnamefont {Katanin}},
  \ and\ \bibinfo {author} {\bibfnamefont {K.}~\bibnamefont {Held}},\ }\href
  {\doibase 10.1103/PhysRevB.75.045118} {\bibfield  {journal} {\bibinfo
  {journal} {Phys. Rev. B}\ }\textbf {\bibinfo {volume} {75}},\ \bibinfo
  {pages} {045118} (\bibinfo {year} {2007})}\BibitemShut {NoStop}%
\bibitem [{\citenamefont {Rohringer}\ \emph {et~al.}(2012)\citenamefont
  {Rohringer}, \citenamefont {Valli},\ and\ \citenamefont
  {Toschi}}]{PhysRevB.86.125114-SM}%
  \BibitemOpen
  \bibfield  {author} {\bibinfo {author} {\bibfnamefont {G.}~\bibnamefont
  {Rohringer}}, \bibinfo {author} {\bibfnamefont {A.}~\bibnamefont {Valli}}, \
  and\ \bibinfo {author} {\bibfnamefont {A.}~\bibnamefont {Toschi}},\ }\href
  {\doibase 10.1103/PhysRevB.86.125114} {\bibfield  {journal} {\bibinfo
  {journal} {Phys. Rev. B}\ }\textbf {\bibinfo {volume} {86}},\ \bibinfo
  {pages} {125114} (\bibinfo {year} {2012})}\BibitemShut {NoStop}%
\bibitem [{\citenamefont {Yin}\ \emph {et~al.}(2014)\citenamefont {Yin},
  \citenamefont {Haule},\ and\ \citenamefont {Kotliar}}]{Nat.Phys.10.845-SM}%
  \BibitemOpen
  \bibfield  {author} {\bibinfo {author} {\bibfnamefont {Z.~P.}\ \bibnamefont
  {Yin}}, \bibinfo {author} {\bibfnamefont {K.}~\bibnamefont {Haule}}, \ and\
  \bibinfo {author} {\bibfnamefont {G.}~\bibnamefont {Kotliar}},\ }\href
  {\doibase 10.1038/nphys3116} {\bibfield  {journal} {\bibinfo  {journal} {Nat.
  Phys.}\ }\textbf {\bibinfo {volume} {10}},\ \bibinfo {pages} {845} (\bibinfo
  {year} {2014})}\BibitemShut {NoStop}%
\bibitem [{\citenamefont {Toschi}\ \emph {et~al.}(2012)\citenamefont {Toschi},
  \citenamefont {Arita}, \citenamefont {Hansmann}, \citenamefont
  {Sangiovanni},\ and\ \citenamefont {Held}}]{PhysRevB.86.064411-SM}%
  \BibitemOpen
  \bibfield  {author} {\bibinfo {author} {\bibfnamefont {A.}~\bibnamefont
  {Toschi}}, \bibinfo {author} {\bibfnamefont {R.}~\bibnamefont {Arita}},
  \bibinfo {author} {\bibfnamefont {P.}~\bibnamefont {Hansmann}}, \bibinfo
  {author} {\bibfnamefont {G.}~\bibnamefont {Sangiovanni}}, \ and\ \bibinfo
  {author} {\bibfnamefont {K.}~\bibnamefont {Held}},\ }\href {\doibase
  10.1103/PhysRevB.86.064411} {\bibfield  {journal} {\bibinfo  {journal} {Phys.
  Rev. B}\ }\textbf {\bibinfo {volume} {86}},\ \bibinfo {pages} {064411}
  (\bibinfo {year} {2012})}\BibitemShut {NoStop}%
\bibitem [{\citenamefont {Hirsch}\ and\ \citenamefont
  {Fye}(1986)}]{PhysRevLett.56.2521-SM}%
  \BibitemOpen
  \bibfield  {author} {\bibinfo {author} {\bibfnamefont {J.~E.}\ \bibnamefont
  {Hirsch}}\ and\ \bibinfo {author} {\bibfnamefont {R.~M.}\ \bibnamefont
  {Fye}},\ }\href {\doibase 10.1103/PhysRevLett.56.2521} {\bibfield  {journal}
  {\bibinfo  {journal} {Phys. Rev. Lett.}\ }\textbf {\bibinfo {volume} {56}},\
  \bibinfo {pages} {2521} (\bibinfo {year} {1986})}\BibitemShut {NoStop}%
\bibitem [{\citenamefont {Gull}\ \emph {et~al.}(2011)\citenamefont {Gull},
  \citenamefont {Millis}, \citenamefont {Lichtenstein}, \citenamefont
  {Rubtsov}, \citenamefont {Troyer},\ and\ \citenamefont
  {Werner}}]{RevModPhys.83.349-SM}%
  \BibitemOpen
  \bibfield  {author} {\bibinfo {author} {\bibfnamefont {E.}~\bibnamefont
  {Gull}}, \bibinfo {author} {\bibfnamefont {A.~J.}\ \bibnamefont {Millis}},
  \bibinfo {author} {\bibfnamefont {A.~I.}\ \bibnamefont {Lichtenstein}},
  \bibinfo {author} {\bibfnamefont {A.~N.}\ \bibnamefont {Rubtsov}}, \bibinfo
  {author} {\bibfnamefont {M.}~\bibnamefont {Troyer}}, \ and\ \bibinfo {author}
  {\bibfnamefont {P.}~\bibnamefont {Werner}},\ }\href {\doibase
  10.1103/RevModPhys.83.349} {\bibfield  {journal} {\bibinfo  {journal} {Rev.
  Mod. Phys.}\ }\textbf {\bibinfo {volume} {83}},\ \bibinfo {pages} {349}
  (\bibinfo {year} {2011})}\BibitemShut {NoStop}%
\bibitem [{\citenamefont {Hafermann}\ \emph {et~al.}(2012)\citenamefont
  {Hafermann}, \citenamefont {Patton},\ and\ \citenamefont
  {Werner}}]{PhysRevB.85.205106-SM}%
  \BibitemOpen
  \bibfield  {author} {\bibinfo {author} {\bibfnamefont {H.}~\bibnamefont
  {Hafermann}}, \bibinfo {author} {\bibfnamefont {K.~R.}\ \bibnamefont
  {Patton}}, \ and\ \bibinfo {author} {\bibfnamefont {P.}~\bibnamefont
  {Werner}},\ }\href {\doibase 10.1103/PhysRevB.85.205106} {\bibfield
  {journal} {\bibinfo  {journal} {Phys. Rev. B}\ }\textbf {\bibinfo {volume}
  {85}},\ \bibinfo {pages} {205106} (\bibinfo {year} {2012})}\BibitemShut
  {NoStop}%
\bibitem [{\citenamefont {Maier}\ \emph {et~al.}(2007)\citenamefont {Maier},
  \citenamefont {Jarrell},\ and\ \citenamefont
  {Scalapino}}]{PhysRevB.75.134519-SM}%
  \BibitemOpen
  \bibfield  {author} {\bibinfo {author} {\bibfnamefont {T.~A.}\ \bibnamefont
  {Maier}}, \bibinfo {author} {\bibfnamefont {M.}~\bibnamefont {Jarrell}}, \
  and\ \bibinfo {author} {\bibfnamefont {D.~J.}\ \bibnamefont {Scalapino}},\
  }\href {\doibase 10.1103/PhysRevB.75.134519} {\bibfield  {journal} {\bibinfo
  {journal} {Phys. Rev. B}\ }\textbf {\bibinfo {volume} {75}},\ \bibinfo
  {pages} {134519} (\bibinfo {year} {2007})}\BibitemShut {NoStop}%
\bibitem [{\citenamefont {Wang}\ \emph {et~al.}(2013)\citenamefont {Wang},
  \citenamefont {Kreisel}, \citenamefont {Zabolotnyy}, \citenamefont
  {Borisenko}, \citenamefont {B\"uchner}, \citenamefont {Maier}, \citenamefont
  {Hirschfeld},\ and\ \citenamefont {Scalapino}}]{PhysRevB.88.174516-SM}%
  \BibitemOpen
  \bibfield  {author} {\bibinfo {author} {\bibfnamefont {Y.}~\bibnamefont
  {Wang}}, \bibinfo {author} {\bibfnamefont {A.}~\bibnamefont {Kreisel}},
  \bibinfo {author} {\bibfnamefont {V.~B.}\ \bibnamefont {Zabolotnyy}},
  \bibinfo {author} {\bibfnamefont {S.~V.}\ \bibnamefont {Borisenko}}, \bibinfo
  {author} {\bibfnamefont {B.}~\bibnamefont {B\"uchner}}, \bibinfo {author}
  {\bibfnamefont {T.~A.}\ \bibnamefont {Maier}}, \bibinfo {author}
  {\bibfnamefont {P.~J.}\ \bibnamefont {Hirschfeld}}, \ and\ \bibinfo {author}
  {\bibfnamefont {D.~J.}\ \bibnamefont {Scalapino}},\ }\href {\doibase
  10.1103/PhysRevB.88.174516} {\bibfield  {journal} {\bibinfo  {journal} {Phys.
  Rev. B}\ }\textbf {\bibinfo {volume} {88}},\ \bibinfo {pages} {174516}
  (\bibinfo {year} {2013})}\BibitemShut {NoStop}%
\bibitem [{\citenamefont {Ahn}\ \emph {et~al.}(2014)\citenamefont {Ahn},
  \citenamefont {Eremin}, \citenamefont {Knolle}, \citenamefont {Zabolotnyy},
  \citenamefont {Borisenko}, \citenamefont {B\"uchner},\ and\ \citenamefont
  {Chubukov}}]{PhysRevB.89.144513-SM}%
  \BibitemOpen
  \bibfield  {author} {\bibinfo {author} {\bibfnamefont {F.}~\bibnamefont
  {Ahn}}, \bibinfo {author} {\bibfnamefont {I.}~\bibnamefont {Eremin}},
  \bibinfo {author} {\bibfnamefont {J.}~\bibnamefont {Knolle}}, \bibinfo
  {author} {\bibfnamefont {V.~B.}\ \bibnamefont {Zabolotnyy}}, \bibinfo
  {author} {\bibfnamefont {S.~V.}\ \bibnamefont {Borisenko}}, \bibinfo {author}
  {\bibfnamefont {B.}~\bibnamefont {B\"uchner}}, \ and\ \bibinfo {author}
  {\bibfnamefont {A.~V.}\ \bibnamefont {Chubukov}},\ }\href {\doibase
  10.1103/PhysRevB.89.144513} {\bibfield  {journal} {\bibinfo  {journal} {Phys.
  Rev. B}\ }\textbf {\bibinfo {volume} {89}},\ \bibinfo {pages} {144513}
  (\bibinfo {year} {2014})}\BibitemShut {NoStop}%
\bibitem [{\citenamefont {Yu}\ \emph {et~al.}(2013)\citenamefont {Yu},
  \citenamefont {Goswami}, \citenamefont {Si}, \citenamefont {Nikolic},\ and\
  \citenamefont {Zhu}}]{Nat.Commun.4.2783-SM}%
  \BibitemOpen
  \bibfield  {author} {\bibinfo {author} {\bibfnamefont {R.}~\bibnamefont
  {Yu}}, \bibinfo {author} {\bibfnamefont {P.}~\bibnamefont {Goswami}},
  \bibinfo {author} {\bibfnamefont {Q.}~\bibnamefont {Si}}, \bibinfo {author}
  {\bibfnamefont {P.}~\bibnamefont {Nikolic}}, \ and\ \bibinfo {author}
  {\bibfnamefont {J.~X.}\ \bibnamefont {Zhu}},\ }\href {\doibase
  10.1038/ncomms3783} {\bibfield  {journal} {\bibinfo  {journal} {Nat Commun}\
  }\textbf {\bibinfo {volume} {4}},\ \bibinfo {pages} {2783} (\bibinfo {year}
  {2013})}\BibitemShut {NoStop}%
\bibitem [{\citenamefont {Hu}\ and\ \citenamefont
  {Ding}(2012)}]{Sci.Rep.2.381-SM}%
  \BibitemOpen
  \bibfield  {author} {\bibinfo {author} {\bibfnamefont {J.}~\bibnamefont
  {Hu}}\ and\ \bibinfo {author} {\bibfnamefont {H.}~\bibnamefont {Ding}},\
  }\href {\doibase 10.1038/srep00381} {\bibfield  {journal} {\bibinfo
  {journal} {Sci. Rep.}\ }\textbf {\bibinfo {volume} {2}},\ \bibinfo {pages}
  {381} (\bibinfo {year} {2012})}\BibitemShut {NoStop}%
\bibitem [{\citenamefont {Hirschfeld}\ \emph {et~al.}(2011)\citenamefont
  {Hirschfeld}, \citenamefont {Korshunov},\ and\ \citenamefont
  {Mazin}}]{0034-4885-74-12-124508-SM}%
  \BibitemOpen
  \bibfield  {author} {\bibinfo {author} {\bibfnamefont {P.~J.}\ \bibnamefont
  {Hirschfeld}}, \bibinfo {author} {\bibfnamefont {M.~M.}\ \bibnamefont
  {Korshunov}}, \ and\ \bibinfo {author} {\bibfnamefont {I.~I.}\ \bibnamefont
  {Mazin}},\ }\href {http://stacks.iop.org/0034-4885/74/i=12/a=124508}
  {\bibfield  {journal} {\bibinfo  {journal} {Reports on Progress in Physics}\
  }\textbf {\bibinfo {volume} {74}},\ \bibinfo {pages} {124508} (\bibinfo
  {year} {2011})}\BibitemShut {NoStop}%
\bibitem [{\citenamefont {Miyake}\ \emph {et~al.}(2010)\citenamefont {Miyake},
  \citenamefont {Nakamura}, \citenamefont {Arita},\ and\ \citenamefont
  {Imada}}]{doi:10.1143/JPSJ.79.044705-SM}%
  \BibitemOpen
  \bibfield  {author} {\bibinfo {author} {\bibfnamefont {T.}~\bibnamefont
  {Miyake}}, \bibinfo {author} {\bibfnamefont {K.}~\bibnamefont {Nakamura}},
  \bibinfo {author} {\bibfnamefont {R.}~\bibnamefont {Arita}}, \ and\ \bibinfo
  {author} {\bibfnamefont {M.}~\bibnamefont {Imada}},\ }\href {\doibase
  10.1143/JPSJ.79.044705} {\bibfield  {journal} {\bibinfo  {journal} {Journal
  of the Physical Society of Japan}\ }\textbf {\bibinfo {volume} {79}},\
  \bibinfo {pages} {044705} (\bibinfo {year} {2010})},\ \Eprint
  {http://arxiv.org/abs/http://journals.jps.jp/doi/pdf/10.1143/JPSJ.79.044705}
  {http://journals.jps.jp/doi/pdf/10.1143/JPSJ.79.044705} \BibitemShut
  {NoStop}%
\bibitem [{\citenamefont {Hajiri}\ \emph {et~al.}(2012)\citenamefont {Hajiri},
  \citenamefont {Ito}, \citenamefont {Niwa}, \citenamefont {Matsunami},
  \citenamefont {Min}, \citenamefont {Kwon},\ and\ \citenamefont
  {Kimura}}]{PhysRevB.85.094509-SM}%
  \BibitemOpen
  \bibfield  {author} {\bibinfo {author} {\bibfnamefont {T.}~\bibnamefont
  {Hajiri}}, \bibinfo {author} {\bibfnamefont {T.}~\bibnamefont {Ito}},
  \bibinfo {author} {\bibfnamefont {R.}~\bibnamefont {Niwa}}, \bibinfo {author}
  {\bibfnamefont {M.}~\bibnamefont {Matsunami}}, \bibinfo {author}
  {\bibfnamefont {B.~H.}\ \bibnamefont {Min}}, \bibinfo {author} {\bibfnamefont
  {Y.~S.}\ \bibnamefont {Kwon}}, \ and\ \bibinfo {author} {\bibfnamefont
  {S.}~\bibnamefont {Kimura}},\ }\href {\doibase 10.1103/PhysRevB.85.094509}
  {\bibfield  {journal} {\bibinfo  {journal} {Phys. Rev. B}\ }\textbf {\bibinfo
  {volume} {85}},\ \bibinfo {pages} {094509} (\bibinfo {year}
  {2012})}\BibitemShut {NoStop}%
\bibitem [{\citenamefont {Nourafkan}\ and\ \citenamefont
  {Tremblay}(2016)}]{1601.05813-SM}%
  \BibitemOpen
  \bibfield  {author} {\bibinfo {author} {\bibfnamefont {R.}~\bibnamefont
  {Nourafkan}}\ and\ \bibinfo {author} {\bibfnamefont {A.-M.}\ \bibnamefont
  {Tremblay}},\ }\href@noop {} {\bibfield  {journal} {\bibinfo  {journal}
  {ArXiv e-prints}\ } (\bibinfo {year} {2016})},\ \Eprint
  {http://arxiv.org/abs/arXiv:1601.05813} {arXiv:1601.05813} \BibitemShut
  {NoStop}%
\end{thebibliography}
\end{document}